\documentclass[aps,pre,reprint,amsmath,amssymb]{revtex4-1}
\usepackage{braket}
\def\withcomments{off}
\def\checkcomments{on}
\ifx\withcomments\checkcomments
        \usepackage{color}
        \usepackage{soul}
        \definecolor{TEcolor}{rgb}{1,0,0}

        \newcommand{\TEout}[1]{\textcolor{TEcolor}{\st{#1}}\marginpar{\textcolor{TEcolor}{\textbar}}}
        \newcommand{\TEcomment}[1]{\textcolor{TEcolor}{{\bf[#1]}}}
        \newcounter{TEmarkcounter}
\else
        \newcommand{\st}[1]{}

        \newcommand{\TEout}[1]{}
        \newcommand{\TEcomment}[1]{}
        \newcounter{TEmarkcounter}
\fi
\begin{document}
\author{Thomas Engl}
\email{Thomas.Engl@physik.uni-regensburg.de}
\affiliation{Institut f\"ur Theoretische Physik, Universit\"at Regensburg, D-93040 Regensburg, Germany}
\author{Juan Diego Urbina}
\affiliation{Institut f\"ur Theoretische Physik, Universit\"at Regensburg, D-93040 Regensburg, Germany}
\author{Klaus Richter}
\affiliation{Institut f\"ur Theoretische Physik, Universit\"at Regensburg, D-93040 Regensburg, Germany}
\title{Periodic Mean-Field Solutions and the Spectra of Discrete Bosonic Fields:\\ Trace Formula for Bose-Hubbard Models}
\begin{abstract}
We consider the many-body spectra of interacting bosonic quantum fields on a lattice in the semiclassical limit of large particle number $N$. We show that the many-body density of states can be expressed as a coherent sum over oscillating long-wavelength contributions given by periodic, non-perturbative solutions of the, typically non-linear, wave equation of the classical (mean-field) limit. To this end we construct the semiclassical approximation for both the smooth and oscillatory part of the many-body density of states in terms of a trace formula starting from the exact path integral form of the propagator between many-body quadrature states. We therefore avoid the use of a complexified classical limit characteristic of the coherent state representation. While quantum effects like vacuum fluctuations and gauge invariance are exactly accounted for, our semiclassical approach captures quantum interference and therefore is valid well beyond the Ehrenfest time where naive quantum-classical correspondence breaks down. Remarkably, due to a special feature of harmonic systems with incommesurable frequencies, our formulas are generically valid also in the free-field case of non-interacting bosons.    
\end{abstract}
\maketitle
%
%
%
%
%
%
%
\section{Introduction}

The full quantum mechanical solution of the problem of interacting particles gets exceedingly complicated with increasing particle number, and even for a generic single-particle problem in the limit of large excitations. Hence there has been the quest for devising versions of the quantum formalism where classical input can be used to predict the outcomes of observations keeping intact concepts like superposition of states and summing amplitudes instead of probabilities as embodied in the kinematical structure of quantum mechanics. A natural benchmark for the use of classical objects in quantum mechanics is the ubiquitous and defining presence of interference phenomena in the quantum world. 

One attempt to search for quantum effects using only classical information consists of following the time evolution of quasiclassical, coherent quantum states with the sharpest distribution of momentum and position allowed by quantum mechanics. The time evolution of minimal wavepackets is then approximated for short times by a rigid motion along the unique classical trajectory fixed by the initial expectation values of position and momentum $\langle \hat{q}(0) \rangle, \langle \hat{p}(0) \rangle$. For times shorter than a usually short characteristic quantum scale, the Ehrenfest time, expectation values are given simply by the classical values $q(t),p(t)$ determined by the unique solution of the classical equations of motion. This approach breaks down when the time evolved expectation values of $\langle \hat{q}(t) \rangle, \langle \hat{p}(t) \rangle$ are insufficient to recover even approximately the time evolved wavepacket. This happens when different sectors of the originally well localized wavepacket start superimposing with each other and produce interference patterns.

The failure of this approach lies in its direct use of classical concepts, as there is no simple way to modify classical mechanics in order to account for interference phenomena, particularly if the approach is fixing a unique classical trajectory. The early realization that quantum phenomena can indeed be explained in terms of interfering amplitudes between classical paths instead of the classical trajectories themselves marks the beginning of the semiclassical program (see \cite{Gutbook} for a historical review). It took, however, almost fifty years until Gutzwiller provided a complete and rigorous derivation of the semiclassical approximation to the quantum mechanical propagator, the starting point of modern semiclassical methods \cite{Gutzwiller_propagator}.

For first-quantized, single-particle systems, Gutzwiller's result for the quantum mechanical density of states (DOS) \cite{Gutzwiller_trace},
\begin{equation}
\rho(E)=\sum_{n}\delta(E-E_{n}),
\end{equation}
where $E_{n}$ are the eigenvalues of the Hamiltonian, has the generic form
\begin{equation}
\rho(E)\simeq \bar{\rho}(E)+\tilde{\rho}(E)
\label{eq:dos_splitting}
\end{equation}
in the formal limit $\hbar \to 0$. Here, the smooth part $\bar{\rho}(E)$ is purely classical, in that it is related with the phase space volume of the classical energy shell, also known as the Weyl term. Remarkably, quantum fluctuations responsible for the oscillatory part $\tilde{\rho}(E)$ are also given in terms of classical quantities, though encoded in a subtle way in the periodic solutions of the classical equations of motion of the corresponding classical system. A key distinction between these two contributions to the semiclassical density of states is that while $\bar{\rho}(E)$ is analytical in $\hbar$ and therefore admits a power expansion around $\hbar=0$, this is not the case for the oscillatory contribution, namely, $\tilde{\rho}(E)$ cannot be approximated by any finite-order expansion in~$\hbar$.

It is natural to ask which modifications, technical or conceptual, are required to take the semiclassical program into the realm of many-body systems where correlation effects due to both interactions and indistinguishability additionally appear. Here we face a unique aspect of many-body systems, namely, that one can choose between two, equivalent but conceptually quite different approaches. On the one hand, one can generalize first-quantized techniques to many-body systems by extending the number of degrees of freedom and using projector techniques to select the states with the appropriate symmetry under permutation. The associated semiclassical approach is then based on interfering classical paths in a multidimensional space, supplemented by boundary conditions and/or extra classical paths joining initial classical configurations with the ones obtained under permutations of the particle labels. This first-quantized approach has been successfully applied to the helium atom, a prototype of a strongly interacting few-particle problem \cite{helium}, and has been used so far to derive formal results for the symmetry-projected trace formula \cite{Weidenmuller,Whelan}. Moreover, the simplifying assumption of a unique mean-field potential fixing the classical dynamics and making it essentially non-interacting has lead to the notion of shell effects and its semiclassical interpretation \cite{shelleffects,Brackbook}. Furthermore, a complementary semiclassical approach to the smooth part of the DOS in many-body systems has shown to be surprisingly accurate \cite{Q} and allows for going beyond the independent particle model.

The above mentioned approaches have in common that the semiclassical limit of high energies ($\hbar \to 0$) is taken for fixed, though possibly large total particle number $N$. The second option to describe quantum many-body systems of interacting identical particles is the use of quantum fields. In this approach, the quantum dynamics for $N\to \infty$ has as classical limit a, typically non-linear, wave equation. This new feature is, however, compensated by the fact that with the notion of quantum fields, indistinguishability is included by construction into the kinematics of the state space, instead of by applying projectors as in the first-quantized approach. The use of quantum fields to describe systems of interacting, identical particles has another key consequence. It has been shown (and it will be apparent from the calculations presented here) that for systems with fixed, finite-dimensional Hilbert space, the classical limit is actually equivalent to $N\to \infty$. It is then reassuring that, as shown in \cite{cbs_fock,my_phd_book}, the classical limit of the second-quantized theory turns out to be the mean-field description, which is expected to correctly represent the dynamics in the thermodynamic limit.    

In this paper, we follow Gutzwiller's program and derive rigorously a formula providing the DOS of a second-quantized system for large $N$ where the classical limit involves discrete field equations, the mean-field equations of the associated discrete quantum field. We construct both the smooth and oscillatory contributions to the quantum many-body DOS starting with the semiclassical approximation to the exact Feynman propagator for the quantum field. We show that the many-body DOS arises from interference of, in principle, infinitely many, periodic solutions of the corresponding classical mean field equations, in close analogy to the periodic orbit contribution to the single-particle DOS. In the case where interactions of strength $g$ are present, our derivation relies on the existence of chaotic behavior in the classical limit (the ubiquitous presence of chaotic regions in the phase space of discrete mean field equations has been addressed in \cite{Sandro}). For this, a suitable scaling of the interaction strength $g\sim N^{-1}$ is usually assumed that keeps its contribution to the total energy $U\sim gN^{2}$ proportional to $N$ and therefore comparable with the single-particle (hopping) term. In the non-interacting case our derivation is valid for systems where the single-particle spectrum used to construct the many-body state space is such that the energies are non-commensurable, as it is generically the case. Our work for the free, non-interacting case opens a road to study in a systematic way the combined limit $N\to \infty, \hbar \to 0$ in infinite-dimensional systems where new kinds of classical structures might become relevant \cite{boris}.    

A central aspect of Gutzwiller's method \cite{Gutzwiller_propagator} is a clever choice of the representation where the semiclassical propagator, the key object representing quantum evolution in terms of solutions of the classical equations of motion, appears as a sum of oscillatory terms given by real actions. While in the first-quantized scenario this choice is naturally given by the position representation, this important aspect of the semiclassical program has not been addressed in the context of quantum fields, where the usual choice for constructing the path integral is the coherent state representation for which the actions entering the semiclassical propagator are complex \cite{Orland}. Thus, here we use a different approach. 

The paper is organized as follows. A key point in our approach is to generalize the concept of position eigenstates into the realm of quantum fields, and for this reason in Sec.~\ref{sec:QMB} we briefly introduce these objects, together with basic definitions of Fock space and creation and annihilation operators. The main technical part of the presentation, Sec.~\ref{sec:DMBDOS}, is dedicated to the rigorous derivation of the many-body DOS for Bose-Hubbard systems starting from its quantum mechanical definition and the semiclassical approximation of the quantum mechanical propagator. Specifically, the smooth (Weyl) contribution to the DOS is derived in Sec.~\ref{sec:MBDOSSP} with its final form given in Eq.~(\ref{eq:Weyl-term}), while the oscillatory, Gutzwiller-like contribution is derived in Sec.~\ref{subsec:osc} and summarized in Eqns.~(\ref{eq:trace-formula}-\ref{eq:Maslov_orig}). The all important derivation of the correct phases accounting for the existence of focal points, the Maslov indeces, is outlined in Sec.~\ref{subsec:maslov}. We finish this technical part with Sec.~\ref{subsec:RedExt} where we explain why the extended phase-space approach used here to account for the peculiar consequences of number conserving Hamiltonians has advantages over a reduced-phase space approach, although the later is more frequently used to deal with continuous symmetries in first-quantized systems. As a first application of our formalism, in Sec.~\ref{sec:FF} we derive a semiclassical expression in terms of intefering periodic mean-field solutions that gives the \emph{exact} form of the many-body DOS for a non-interacting bosonic field with incommensurable single-particle energies. The connection of our approach with some existing semiclassically inspired ideas and methods in the context of interacting bosonic systems is presented in Sec.~\ref{sec:EXT}, together with possible applications. Finally, in Sec.~\ref{sec:CONC} we conclude with a summary of our findings and some of its consequences.    

\section{Quantum mechanical background}
\label{sec:QMB}
We will restrict ourselves to quantum fields described by a general Bose-Hubbard Hamiltonian with two-body interactions,
\begin{equation}
\hat{H}=\sum\limits_{l_1,l_2=1}^{L}H_{l_1l_2}\hat{a}_{l_1}^\dagger\hat{a}_{l_2}+\frac{1}{2}\sum\limits_{l_1,l_2,l_3,l_4=1}^{L}U_{l_1l_2l_3l_4}\hat{a}_{l_1}^\dagger\hat{a}_{l_2}^\dagger\hat{a}_{l_3}\hat{a}_{l_4}.
\label{eq:QH}
\end{equation}
Here, ${\bf H}=\left(H_{l_1l_2}\right)_{l_1,l_2=1,\ldots,L}$ is the hermitian matrix describing the single-particle motion and the four-fold sum describes two-body interactions. Moreover, $\hat{a}_l$ and $\hat{a}_l^\dagger$ are the annihilation and creation operators for the $l$-th single-particle state (or site) satisfying the usual bosonic commutation relations $\left[\hat{a}_l,\hat{a}_{l^\prime}^{\dagger}\right]=\delta_{ll^\prime}$.

At intermediate steps, we will make use of Fock states $\ket{\bf n}$ determined by the (integer) occupation numbers $n_1,\ldots,n_L$. These states satisfy $\hat{a}_l^\dagger\hat{a}_l\ket{\bf n}=n_l\ket{\bf n}$ and
\begin{equation}
\begin{split}
\ket{\bf n}=&\frac{1}{\sqrt{\prod\limits_{l=1}^{L}n_l!}}\left(\hat{a}_L^\dagger\right)^{n_L}\cdots\left(\hat{a}_1^\dagger\right)^{n_1}\ket{\bf 0}, 
\end{split}
\end{equation} 
More important for the derivation of the trace formula, however, are so-called quadrature eigenstates $\ket{{\bf q}}$ and $\ket{\bf p}$ defined by the eigenvalue equations \cite{VogelWelsch}
\begin{equation}
\begin{split}
\frac{1}{2}\left(\hat{a}_l^{}+\hat{a_l}^\dagger\right)\ket{\bf q}&=q_l\ket{\bf q}, \\
-\frac{\rm i}{2}\left(\hat{a}_l^{}-\hat{a_l}^\dagger\right)\ket{\bf p}&=p_l\ket{\bf p},
\end{split}
\end{equation}
which satisfy
\begin{equation}
\begin{split}
\Braket{{\bf q}|{\bf n}}=&\prod\limits_{l=1}^{L}\frac{{\rm e}^{-q_l^2}}{\sqrt{2^{n-1}n!\sqrt{2\pi}}}H_{n_l}\left(\sqrt{2}q_l\right),  
\end{split}
\end{equation}
and
\begin{equation}
\begin{split}
\Braket{{\bf p}|{\bf n}}=&\prod\limits_{l=1}^{L}\frac{{\rm e}^{-p_l^2+{\rm i}n\frac{\pi}{2}}}{\sqrt{2^{n-1}n!\sqrt{2\pi}}}H_{n_l}\left(\sqrt{2}p_l\right), 
\end{split}
\end{equation}
where $H_n$ denotes the $n$-th Hermite polynomial. We have also the overlap 
\begin{equation}
\begin{split}
\Braket{{\bf q}|{\bf p}}=&\prod\limits_{l=1}^{L}\frac{{\rm e}^{2{\rm i}p_lq_l}}{\sqrt{\pi}},
\end{split}
\end{equation}
and the closure relations
\begin{equation}
\begin{split}
\hat{1}=&\int{\rm d}^Lq\ket{\bf q}\bra{\bf q}=\int{\rm d}^Lp\ket{\bf p}\bra{\bf p}.
\end{split}
\end{equation}

For the derivation of the smooth part of the DOS, we will later make use of the asymptotic formula of the Hermite polynomials for large $n$ \cite{Brackbook},
\begin{widetext}
\begin{equation}
\Braket{{\bf q}|{\bf n}}=\prod\limits_{l=1}^{L}\frac{\cos\left\{q_l\sqrt{\left(n_l+\frac{1}{2}\right)-q_l^2}-\left(n_l+\frac{1}{2}\right)\arccos\left(\frac{q_l}{\sqrt{n_l+\frac{1}{2}}}\right)\right\}}{\sqrt{\frac{\pi}{2}}\sqrt[4]{\left(n_l+\frac{1}{2}\right)-q_l^2}}.
\label{eq:asymptotic_hermite}
\end{equation}
\end{widetext}
\section{Derivation of the many-body density of states}
\label{sec:DMBDOS} 
The DOS $\rho_N(E)$ for fixed number of particles $N$ is given by the imaginary part of the trace of the Green function $\hat{G}(E)$ over the subspace of the full Hilbert space obtained by fixing $N=\sum_{l=1}^{L}n_l$. It is given by
\begin{equation}
\rho_N\left(E\right)=-\frac{1}{\pi}\lim\limits_{\eta\to0}\Im g_N(E+{\rm i}\eta),
\label{eq:dos_definition}
\end{equation}
with $\Im$ denoting the imaginary part and with the trace in terms of a sum over Fock states
\begin{equation}
g_N(E)={\rm Tr}_N\hat{G}\left(E\right)=\sum\limits_{\bf n}\delta_{\sum\limits_{l=1}^{L}n_{l},N}\Braket{\bf n|\hat{G}\left(E\right)|\bf n}.
\label{eq:connection_dos_green}
\end{equation}
Semiclassically, the single-particle DOS is typically split up into a smooth part, which stems from short trajectories and an oscillatory part determined by periodic orbits, see Eq.~(\ref{eq:dos_splitting}). Correspondingly, for the many-body case we will now first derive the smooth part $\bar{\rho}_N(E)$.
\subsection{Smooth part}
\label{sec:MBDOSSP}
To this end we first rewrite the sum over all possible occupations and the Kronecker delta  in Eq.~(\ref{eq:connection_dos_green}) by a sum over those occupations, which have the correct total number of particles and insert the definition of the Green function as a Laplace transform of the propagator $\hat{K}(t)$ in Fock space,
\begin{equation}
\begin{split}
g_N(E)&=\sum\limits_{{\bf n}:\sum\limits_{l=1}^{L}n_l=N}\Braket{\bf n|\hat{G}\left(E\right)|\bf n} \\
&=\frac{1}{{\rm i}\hbar}\int\limits_{0}^{\infty}{\rm d}t\,{\rm e}^{\frac{\rm i}{\hbar}Et}\sum\limits_{{\bf n}:\sum\limits_{l=1}^{L}n_l=N}\Braket{\bf n|\hat{K}\left(t\right)|\bf n}.
\end{split}
\label{eq:resolvent_smooth_start}
\end{equation}
The smooth part of the DOS stems from short paths, {\it i.e.}~from the short time contribution to the integral. In order to compute this contribution, we will first evaluate the trace and then perform the integration. To this end we rewrite the diagonal matrix elements as
\begin{align}
&\Braket{\bf n|\hat{K}\left(t\right)|\bf n}={\rm Tr}\hat{K}(t)\ket{\bf n}\bra{\bf n}=
\label{eq:propagator_matrix_element_fock} \\
&\qquad\int{\rm d}^Lq\int{\rm d}^Lp\left[\hat{K}(t)\right]_{\rm Weyl}({\bf q},{\bf p})\left[\ket{{\bf n}}\bra{\bf n}\right]_{\rm Weyl}({\bf q},{\bf p}) \nonumber
\end{align}
with the Weyl symbols of an operator $\hat{O}$ being defined by \cite{OdA}
\begin{align}
&\left[\hat{O}\right]_{\rm Weyl}({\bf q},{\bf p})=
\label{eq:definition_Weyl-Symbol} \\
&\int{\rm d}^LQ\Braket{{\bf q}+\frac{\bf Q}{2}|\hat{O}|{\bf q}-\frac{\bf Q}{2}}\Braket{{\bf q}-\frac{\bf Q}{2}|{\bf p}}\Braket{{\bf p}|{\bf q}+\frac{\bf Q}{2}}. \nonumber
\end{align}

Next, we use the asymptotic formula (\ref{eq:asymptotic_hermite}) for the Hermite polynomials for large $n$ and rewrite the cosines as exponentials, yielding four terms, where for two of them the exponents from the cosines have the same sign, while for the remaining two these signs are different. However, in Eq.~(\ref{eq:definition_Weyl-Symbol}) replacing ${\bf Q}$ by $-{\bf Q}$ is the same as complex conjugation, which is only true, if the two signs of the exponential are opposite. Therefore, the terms with both signs being the same have to cancel when performing the integral. Thus, the resulting exponent is antisymmetric in $Q_l$. Expanding it up to second order in $Q_l$ and neglecting the dependence of the prefactor on $Q_l$ yields
\begin{widetext}
\begin{equation}
\begin{split}
\left[\ket{{\bf n}}\bra{\bf n}\right]_{\rm Weyl}({\bf q},{\bf p})\approx&\prod\limits_{l=1}^{L}\sum\limits_{s_l=\pm1}2\frac{\int\limits_{-\infty}^{\infty}{\rm d}Q_l\exp\left[-2{\rm i}Q_l\left(p_l+s_l\sqrt{\left(n_l+\frac{1}{2}\right)-q_l^2}\right)\right]}{\pi^2\sqrt{\left(n_l+\frac{1}{2}\right)-q_l^2}} \\
=&\prod\limits_{l=1}^{L}\sum\limits_{s_l=\pm1}\frac{2}{\pi}\frac{\delta\left(p_l+s_l\sqrt{\left(n_l+\frac{1}{2}\right)-q_l^2}\right)}{\sqrt{\left(n_l+\frac{1}{2}\right)-q_l^2}} \\
=&\prod\limits_{l=1}^{L}\frac{1}{\pi b^2}\delta\left(n_l+\frac{1}{2}-q_l^2-p_l^2\right).
\end{split}
\label{eq:Weyl_Symbol_density-matrix}
\end{equation}
\end{widetext}

For the Weyl symbol of the propagator, one can use the usual short time asymptotic form
\begin{equation}
\left[\hat{K}(t)\right]_{\rm Weyl}({\bf q},{\bf p})\approx\exp\left[-\frac{\rm i}{\hbar}H^{(MF)}\left({\bf p},{\bf q}\right)t\right].
\label{eq:Weyl_Symbol_propagator}
\end{equation}
Here,
\begin{align}
\label{eq:HMF}
&H^{(MF)}\left({\bf p},{\bf q}\right)=\frac{\Braket{{\bf p}|\hat{H}|{\bf q}}}{\Braket{{\bf p}|{\bf q}}}= \\
&\sum\limits_{l_1,l_2=1}^{L}\left(h_{l_1l_2}-\frac{1}{2}\sum\limits_{l_3=1}^{L}U_{l_1l_3l_3l_2}\right)\left(\psi_{l_1}^\ast\psi_{l_2}-\frac{1}{2}\delta_{l_1l_2}\right) \nonumber \\
&+\frac{1}{2}\sum\limits_{l_1l_2l_3l_4}U_{l_1l_2l_3l_4}\left(\psi_{l_1}^\ast\psi_{l_3}-\frac{1}{2}\delta_{l_1l_3}\right)\left(\psi_{l_2}^\ast\psi_{l_4}-\frac{1}{2}\delta_{l_2l_4}\right) \nonumber
\end{align}
is the mean field Hamiltonian $H^{(MF)}$ corresponding to the full quantum Hamiltonian (\ref{eq:QH}). It can be obtained by the simple replacement rule \cite{cbs_fock,my_phd_book}
\begin{equation}
\hat{a}_{l}^\dagger\hat{a}_{l^{\prime}}\to\psi_l^\ast\psi_{l^\prime}-\frac{1}{2}\delta_{ll^{\prime}},
\label{eq:rep}
\end{equation}
with $\psi_l=q_l+{\rm i}p_l$.

Inserting Eqns.~(\ref{eq:Weyl_Symbol_density-matrix}) and (\ref{eq:Weyl_Symbol_propagator}) into Eq.~(\ref{eq:propagator_matrix_element_fock}) as well as replacing the sum over occupations in Eq.~(\ref{eq:resolvent_smooth_start}) by an integral then yields for the smooth part of the resolvent
\begin{widetext}
\begin{equation}
\bar{g}_N(E)=\frac{1}{{\rm i}\hbar\left(\frac{\pi}{4}\right)^L}\int\limits_{0}^{\infty}{\rm d}t\int{\rm d}^Lq\int{\rm d}^Lp\exp\left\{\frac{\rm i}{\hbar}\left[E-H^{(MF)}\left({\bf p},{\bf q}\right)\right]t\right\}\delta\left({\bf q}^2+{\bf p}^2-N-\frac{L}{2}\right),
\end{equation}
and thus for the smooth part of the many-body DOS
\begin{equation}
\bar{\rho}_N(E)=\left(\frac{4}{\pi}\right)^L\int{\rm d}^Lq\int{\rm d}^Lp\delta\left(E-H^{(MF)}\left({\bf p},{\bf q}\right)\right)\delta\left({\bf q}^2+{\bf p}^2-N-\frac{L}{2}\right).
\label{eq:Weyl-term}
\end{equation}
\end{widetext}
As in the single-particle case, the smooth part is given by the phase space volume of the $N$-particle energy shell.
\subsection{Oscillatory part}
\label{subsec:osc}
To compute the oscillatory part $\tilde{g}_{N}(E)$, we start again from Eq.~(\ref{eq:connection_dos_green}) and rewrite the Kronecker delta, to get the resolvent as
\begin{equation}
g_N(E)=\frac{1}{2\pi}\int\limits_{0}^{2\pi}{\rm d}\alpha\sum\limits_{\bf n}\Braket{\bf n|{\rm e}^{-{\rm i}\alpha\left(N-\sum\limits_{l=1}^{L}\hat{a}_l^\dagger\hat{a}_l^{}\right)}\hat{G}\left(E\right)|\bf n}.
\end{equation}

The oscillatory part of the DOS, which we are interested in here, can be obtained from a semiclassical approximation of the Green function by computing the trace using a stationary phase approximation. However, for Fock space, the stationary phase approximation is not applicable, since the trace in Fock states is given by a sum rather than an integral. On the other hand, in \cite{cbs_fock,my_phd_book} a possible way to circumvent this problem has been shown. This is by again using the quadrature eigenstates $\ket{\bf q},\ket{\bf p}$. Inserting them to the left and to the right of the Green function yields
\begin{align}
g_N\left(E\right)=\rule{6cm}{0pt}
\label{eq:dos_start} \\
\frac{1}{2\pi}\int\limits_{0}^{2\pi}{\rm d}\alpha\int\limits_{\infty}^{\infty}{\rm d}^Lq\int\limits_{\infty}^{\infty}{\rm d}^Lp\sum\limits_{\bf n}\Braket{{\bf p}|\hat{G}\left(E\right)|{\bf q}}\Braket{{\bf q}|{\bf n}} \nonumber \\
\times\Braket{\bf n|{\rm e}^{-{\rm i}\alpha\left(N-\sum\limits_{l=1}^{L}\hat{a}_l^\dagger\hat{a}_l^{}\right)}|\bf p}&. \nonumber
\end{align}
Using the completeness relation of the Hermite polynomials then yields
\begin{equation}
\begin{split}
&\sum\limits_{\bf n}\Braket{{\bf q}|{\bf n}}\Braket{\bf n|{\rm e}^{{\rm i}\alpha\sum\limits_{l=1}^{L}\hat{a}_l^\dagger\hat{a}_l^{}}|\bf p} \\
&=\prod\limits_{l=1}^{L}\frac{\exp\left\{\frac{\rm i}{\cos\alpha}\left[2q_lp_l+\left(q_l^2+p_l^2\right)\sin\alpha\right]-{\rm i}\frac{\alpha}{2}\right\}}{\sqrt{\pi\cos\alpha}}.
\end{split}
\label{eq:particle-number-fixing-term}
\end{equation}

Next, one has to find an expression for the Green function, which is related to the propagator by means of a Laplace transform,
\begin{equation}
G\left({\bf p},{\bf q};E\right)=\Braket{{\bf p}|\hat{G}\left(E\right)|{\bf q}}
=\frac{1}{{\rm i}\hbar}\int\limits_{0}^{\infty}{\rm d}t{\rm e}^{\frac{\rm i}{\hbar}Et}K\left({\bf p},{\bf q},t\right).
\label{eq:connection_green_propagator}
\end{equation}
In \cite{my_phd_book}, a semiclassical approximation for the propagator has been found, which is given by
\begin{equation}
K\left({\bf p},{\bf q},t\right)=\sum\limits_\gamma\sqrt{\left|\det\frac{1}{2\pi\hbar}\frac{\partial^2R_\gamma}{\partial{\bf p}\partial{\bf q}}\right|}{\rm e}^{\frac{\rm i}{\hbar}R_\gamma-{\rm i}\tilde{\mu}_{\gamma}\frac{\pi}{2}}
\end{equation}
where the sum runs over all mean-field trajectories (nonlinear waves) $\gamma$ given by the solutions of the equations of motion
\begin{equation}
{\rm i}\hbar\dot{\boldsymbol \psi}(t)=\frac{\partial H^{({\rm MF})}\left({\boldsymbol \psi}^\ast(t),{\boldsymbol \psi}(t)\right)}{\partial{\boldsymbol \psi}^\ast(t)},
\label{eq:eom}
\end{equation}
and the boundary conditions
\begin{equation}
\begin{split}
\Re{\boldsymbol \psi}(0)&={\bf q}, \\
\Im{\boldsymbol \psi}(t)&={\bf p}.
\end{split}
\label{eq:bc}
\end{equation}
Moreover, the phase each trajectory contributes with is given by its action
\begin{equation}
\begin{split}
R_\gamma=&\int\limits_0^t{\rm d}t^\prime\left[2\hbar\Im{\boldsymbol \psi}(t^\prime)\cdot\Re\dot{\boldsymbol \psi}(t^\prime)-H^{({\rm MF})}\left({\boldsymbol \psi}^\ast(t^\prime),{\boldsymbol \psi}(t^\prime)\right)\right] \\
&-2{\bf p}\cdot\Re{\boldsymbol \psi}(t)
\end{split}
\end{equation}
and the Morse index $\tilde{\mu}_{\gamma}$.

For later reference, we state the derivatives of the action with respect to $\bf p$, $\bf q$ and $t$:
\begin{equation}
\begin{split}
\frac{\partial R_\gamma}{\partial \bf p}=&-2\hbar\Re{\boldsymbol \psi}(0), \\
\frac{\partial R_\gamma}{\partial \bf q}=&-2\hbar\Im{\boldsymbol \psi}(t), \\
\frac{\partial R_\gamma}{\partial t}=&-H^{({\rm MF})}\left({\boldsymbol \psi}^\ast(0),{\boldsymbol \psi}(0)\right)=-E_\gamma.
\end{split}
\end{equation}
In order to determine the oscillatory part of the many-body DOS, the time integration in Eq.~ (\ref{eq:connection_green_propagator}) can be evaluated using a stationary phase approximation. The stationarity condition then selects those trajectories which have energy $E$,
\begin{equation}
\frac{\partial}{\partial t}\left[R_\gamma+Et\right]=E-E_\gamma=0
\end{equation}
In order to compute the semiclassical prefactor of the Green function, one can use the standard trick for Jacobians, with $T_{\gamma}$ the period of $\gamma$, \cite{ChaosBook33}
\begin{eqnarray}
\begin{split}
\det\frac{\partial\left(\Im{\boldsymbol \psi}(0),T_\gamma\right)}{\partial\left({\bf p},E\right)}&=\det\left(\frac{\partial\left(\Im{\boldsymbol \psi}(0),T_\gamma\right)}{\partial\left({\bf p},T_\gamma\right)}\frac{\partial\left({\bf p},T_\gamma\right)}{\partial\left({\bf p},E\right)}\right) \\
&=\det\left(\frac{\partial\Im{\boldsymbol \psi}(0)}{\partial{\bf p}}\right)\frac{\partial T_\gamma}{\partial E}.
\end{split}
\end{eqnarray}
With this, the semiclassical Green function Eq.~(\ref{eq:connection_green_propagator}) is given by
\begin{align}
G&\left({\bf p},{\bf q};E\right)=
\label{eq:greens_semicl_nomaslov} \\
&\frac{1}{{\rm i}\hbar}\frac{1}{\sqrt{2\pi\hbar}^{L-1}}\sum\limits_{\gamma}\sqrt{\left|\det\left(
\begin{array}{cc}
\frac{\partial^2W_\gamma}{\partial{\bf q}\partial{\bf p}} & \frac{\partial^2W_\gamma}{\partial{\bf q}\partial E} \\
\frac{\partial^2W_\gamma}{\partial E\partial{\bf p}} & \frac{\partial^2W_\gamma}{\partial E^2}
\end{array}
\right)\right|}{\rm e}^{\frac{\rm i}{\hbar}W_\gamma-{\rm i}\mu_{\gamma}\frac{\pi}{2}}, \nonumber
\end{align}
with $\mu_\gamma=\tilde{\mu}_\gamma+{\rm sign}(\partial E/\partial t_\gamma)/2$ and
\begin{equation}
W_\gamma=R_\gamma+ET_\gamma=2\hbar\int\limits_{0}^{T_\gamma}\Im{\boldsymbol \psi(t)}\cdot\Re\dot{\boldsymbol \psi}(t){\rm d}t-2\hbar{\bf p}\cdot\Re{\boldsymbol \psi}(T_\gamma)
\end{equation}
satisfying
\begin{equation}
\begin{split}
\frac{\partial W_\gamma}{\partial \bf p}=&-2\hbar\Re{\boldsymbol \psi}(0), \\
\frac{\partial W_\gamma}{\partial \bf q}=&-2\hbar\Im{\boldsymbol \psi}(T_{\gamma}), \\
\frac{\partial W_\gamma}{\partial E}=&T_\gamma.
\end{split}
\end{equation}
Thus, in the semiclassical limit the oscillatory contribution of the resolvent (\ref{eq:dos_start}) is given by
\begin{widetext}
\begin{equation}
\tilde{g}_N(E)=
\frac{1}{2\pi{\rm i}\hbar}\int\limits_{0}^{2\pi}{\rm d}\alpha{\rm e}^{-{\rm i}\alpha N}\int{\rm d}^Lq\int{\rm d}^Lp\sum\limits_{\gamma}\sqrt{\left|\det\left(
\begin{array}{cc}
\frac{\partial^2W_\gamma}{\partial{\bf q}\partial{\bf p}} & \frac{\partial^2W_\gamma}{\partial{\bf q}\partial E} \\
\frac{\partial^2W_\gamma}{\partial E\partial{\bf p}} & \frac{\partial^2W_\gamma}{\partial E^2}
\end{array}
\right)\right|}\frac{{\rm e}^{\frac{\rm i}{\hbar}W_\gamma-{\rm i}\frac{L}{2}\alpha-{\rm i}\mu_{\gamma}\frac{\pi}{2}+\frac{{\rm i}\left[2{\bf q}\cdot{\bf p}+\left({\bf q}^2+{\bf p}^2\right)\sin\alpha\right]}{\cos\alpha}}}{\sqrt{2\pi\hbar}^{L-1}\sqrt{\pi\cos\alpha}^L}.
\label{eq:dos_before_spa}
\end{equation}
\end{widetext}
The integrations over $\bf p$ and $\bf q$ as well as $\alpha$ will again be performed in stationary phase approximation.

The corresponding stationary phase conditions for the integrations over $\bf p$ and $\bf q$ read
\begin{widetext}
\begin{align}
\frac{\partial}{\partial\bf q}\left\{W_\gamma+\frac{\left[2{\bf q}\cdot{\bf p}+\left({\bf q}^2+{\bf p}^2\right)\sin\alpha\right]}{\cos\alpha}\right\}=-2\left[\Im{\boldsymbol \psi}(0)-\frac{{\bf q}\sin\alpha}{\cos\alpha}-\frac{\bf p}{\cos\alpha}\right]=0,
\label{subeq:spc_q} \\
\frac{\partial}{\partial\bf p}\left\{W_\gamma+\frac{\left[2{\bf q}\cdot{\bf p}+\left({\bf q}^2+{\bf p}^2\right)\sin\alpha\right]}{\cos\alpha}\right\}=-2\left[\Re{\boldsymbol \psi}(T_\gamma)-\frac{{\bf p}\sin\alpha}{\cos\alpha}-\frac{\bf q}{\cos\alpha}\right]=0,
\label{subeq:spc_p}
\end{align}
\end{widetext}
which can be combined into the more compact condition
\begin{equation}
{\boldsymbol \psi}(T_\gamma)={\boldsymbol \psi}(0){\rm e}^{-{\rm i}\alpha}.
\label{eq:pseudoperidicity}
\end{equation}
Equation (\ref{eq:pseudoperidicity}) implies that the resulting trace formula will be given by a sum over \emph{pseudo-}periodic orbits, \footnote{Note that the expression pseudo-orbit has been used in a related but different context, there referring to the multilinear combinations of orbits that appear in the semiclassical expression of the spectral zeta function \cite{Keat-Berry}. Here we use pseudo-periodicity as defined the in Eq.~(\ref{eq:pseudoperidicity}).} for which the associated classical nonlinear waves after a certain pseudo-period $T_\gamma$ differ from their initial values by a global phase $\alpha$.

Representing the classical nonlinear wave solution in terms of its amplitude and phase,
\begin{equation}
\psi_l(t)=\sqrt{n_l(t)}{\rm e}^{{\rm i}\theta_l(t)},
\end{equation}
the resulting stationary phase solution for the $\alpha$-integration in Eq.~(\ref{eq:dos_before_spa}) is given by
\begin{equation}
\begin{split}
\tilde{S}_{\gamma}&=W_\gamma+\frac{\left[2{\bf q}\cdot{\bf p}+\left({\bf q}^2+{\bf p}^2\right)\sin\alpha\right]}{\cos\alpha} \\
&=\hbar\int\limits_{0}^{T_{\gamma}}{\boldsymbol\theta}(t)\cdot\dot{\bf n}(t){\rm d}t+\hbar {\bf n}(0)\cdot\left[{\boldsymbol\theta}(0)-{\boldsymbol\theta}(T_{\gamma})\right].
\end{split}
\label{eq:action_fixed_alpha}
\end{equation}
In the last term, which originates from a partial integration, ${\bf n}(T_{\gamma})={\bf n}(0)$ has been used. Its dependence on $\alpha$ is determined by
\begin{equation}
{\bf n}(0)\cdot\left[{\boldsymbol\theta}(0)-{\boldsymbol\theta}(T_{\gamma})\right]=N_{\gamma}\alpha+2\pi\sum\limits_{l=1}^{L}n_l(0) k_l,
\end{equation}
with
\begin{equation}
\label{eq:action_fixed_alphaJD}
N_{\gamma}=\sum\limits_{l=1}^{l}\left|\psi_l(0)\right|^2
\end{equation}
the (time independent) number of particles defined by the trajectory and $k_1,\ldots,k_L$ being integers.

Taking a closer look, one recognizes that due to the conservation of the total number of particles already $2L-1$ of the stationary phase conditions, Eqns.~(\ref{subeq:spc_q},\ref{subeq:spc_p}) suffice to satisfy all of them. Due to the Noether theorem, there is a continuous symmetry for each conserved quantity. Here, this continuous symmetry is given by the $U(1)$ gauge symmetry, \textit{i.e.}~the freedom to chose an arbitrary time-independent global phase $\theta$.

Moreover, as in Gutzwiller's original derivation \cite{Gutzwiller_trace} of the single-particle trace formula, the starting point of the pseudo-periodic orbit can be chosen at any point along the orbit. Thus, there remain two integrations, which are the integrations over all pseudo-periodic orbits belonging to the same continuous family of trajectories, that have to be performed exactly.

For single-particle systems, the trace formula for chaotic systems with additional continuous symmetries has been studied in \cite{traceformula_cont_sym}. The evaluation of the semiclassical prefactor of the trace formula presented there can, to a large extend, be carried over straight forwardly with minor modifications in order to correctly account for the fact that the orbits in the case studied here are not strictly periodic. Therefore, here we will only show the steps, which have to be altered and refer the reader to Ref.~\cite{traceformula_cont_sym} for more details.

After transforming in Eq.~(\ref{eq:dos_before_spa}) the integration variables ${\bf q},{\bf p}$ locally to ${\bf q}_{\parallel},{\bf q}_{\perp},{\bf p}_{\parallel},{\bf p}_{\perp}$, where the (two-dimensional) parallel components run along directions of the continuous families, {\it i.e.}~along the trajectory and the direction of the global phase $\theta$, while the remaining ones are perpendicular to these, the integrations over the perpendicular components as well as ${\bf p}_\parallel$ yield
\begin{widetext}
\begin{equation}
\tilde{g}_N(E)=\frac{1}{2\pi}\int\limits_{0}^{2\pi}{\rm d}\alpha\int{\rm d}^2q_{\parallel}\sum\limits_{\gamma}\frac{{\rm e}^{-{\rm i}\alpha\left(N+\frac{L}{2}\right)}}{{\rm i}\sqrt{2\pi\hbar}\sqrt{\cos\alpha}^L}\left|\det\frac{\partial\left(\Im{\boldsymbol\psi}(0),T_\gamma\right)}{\partial\left({\bf p},E\right)}\det\left(\begin{array}{cc}
\frac{\partial\Im{\boldsymbol\psi}_{\perp}(0)}{\partial{\bf q}_{\perp}}-\frac{\sin\alpha}{\cos\alpha} & \frac{\partial\Im{\boldsymbol\psi}_{\perp}(0)}{\partial\bf p}-\frac{1}{\cos\alpha} \\
\frac{\partial\Re{\boldsymbol\psi}(T_{\gamma})}{\partial{\bf q}_{\perp}}-\frac{1}{\cos\alpha} & \frac{\partial\Re{\boldsymbol\psi}(T_{\gamma})}{\partial{\bf p}}-\frac{\sin\alpha}{\cos\alpha}
\end{array}\right)^{-1}\right|^{\frac{1}{2}}{\rm e}^{\frac{\rm i}{\hbar}\tilde{S}_\gamma-{\rm i}\frac{\pi}{2}\left(\mu_{\gamma}+\nu_{\gamma}\right)},
\label{eq:dos_directly_after_perpendicular_integrations}
\end{equation}
\end{widetext}
where $\nu_{\gamma}=(N_{+}-N_{-})/2$ is the difference between the number of positive and negative eigenvalues of the $(2L-2)\times(2L-2)$ dimensional matrix appearing in the semiclassical prefactor. Note that the sum runs over pseudo-periodic orbits with the initial global phase and the initial position within the orbit chosen by the integration values. Alternatively, one can also refer to the sum over $\gamma$ as a sum over \emph{families} of pseudo-periodic orbits, where one is free to choose the initial global phase of the reference orbit, which is used to compute its contribution.

Leaving the calculation of the determinant in Eq.~(\ref{eq:dos_directly_after_perpendicular_integrations}) to Appendix \ref{app:trace-prefactor}, the trace of the semiclassical Greens function is given by
\begin{widetext}
\begin{equation}
\tilde{g}_N(E)=\frac{1}{2\pi}\int\limits_{0}^{2\pi}{\rm d}\alpha\int{\rm d}^2q_{\parallel}\sum\limits_{\gamma}\frac{{\rm e}^{-{\rm i}\alpha\left(N+\frac{L}{2}\right)}}{{\rm i}\hbar\sqrt{2\pi}}\frac{1}{\sqrt{\left|\det\left(M_{\gamma}-1\right)\frac{\partial\theta}{\partial N_{\gamma}}\right|}}\left|\frac{\partial\Re\left({\boldsymbol\psi}_{\parallel}\left(T_{\gamma}\right){\rm e}^{{\rm i}\alpha}\right)}{\partial\left(T_{\gamma},\theta\right)}\right|^{-1}{\rm e}^{\frac{\rm i}{\hbar}\tilde{S}_{\gamma}-{\rm i}\frac{\pi}{2}\left(\mu_{\gamma}+\nu_{\gamma}+L\eta(\alpha)\right)}.
\label{eq:dos_before_parallel_integrations}
\end{equation}
\end{widetext}
Here $\theta$ is the initial phase of the trajectory, 
\begin{equation}
\eta(\alpha)=\begin{cases}
1 & \text{if }\frac{\pi}{2}<\alpha<\frac{3\pi}{2}, \\
0 & \text{else}
\end{cases}
\end{equation}
and
\begin{equation}
\begin{split}
M_{\gamma}=&\frac{\partial\left(\Re\left({\boldsymbol\psi}_{\perp}(T_{\gamma}){\rm e}^{{\rm i}\alpha}\right),{\bf p}_{\perp}\right)}{\partial\left({\bf q}_{\perp},\Im\left({\boldsymbol\psi}_{\perp}\left(0\right){\rm e}^{-{\rm i}\alpha}\right)\right)} \\
=&\frac{\partial\left(\Re\left({\boldsymbol\psi}_{\perp}(T_{\gamma}){\rm e}^{{\rm i}\alpha}\right),\Im\left({\boldsymbol \psi}_{\perp}(T_{\gamma}){\rm e}^{{\rm i}\alpha}\right)\right)}{\partial\left(\Re\left({\boldsymbol \psi}_{\perp}\left(0\right)\right),\Im\left({\boldsymbol\psi}_{\perp}\left(0\right)\right)\right)}
\label{eq:stability-matrix}
\end{split}
\end{equation}
is the stability matrix for the pseudo-periodic orbit. In view of Eqs.~(\ref{eq:bc},\ref{eq:pseudoperidicity}), for the pseudo-periodic orbit
\begin{equation}
\begin{split}
\Im\left({\boldsymbol\psi}(0){\rm e}^{-{\rm i}\alpha}\right)&={\bf p}, \\
\Re\left({\boldsymbol\psi}(T_{\gamma}){\rm e}^{{\rm i}\alpha}\right)&={\bf q}
\end{split}
\label{eq:pseudoperiodicity_orig_with_pq}
\end{equation}
holds.

Hence, the matrix in Eq.~(\ref{eq:stability-matrix}) is indeed the many-body, field-theoretic analogue to the monodromy matrix appearing in the usual Gutzwiller trace formula \cite{Gutzwiller_trace}.

Now, in Eq.~(\ref{eq:dos_before_parallel_integrations}), the last determinant can be used in order to transform the integration over ${\bf q}_{\parallel}$ into integrations over the propagation time and the global phase. Again in view of Gutzwiller's derivation \cite{Gutzwiller_trace} one has to correctly account for repetitions of each primitive pseudo-periodic orbit when evaluating these integrals. These primitive pseudo-periodic orbits are obtained by finding the largest possible, but finite integer $m\geq1$ for which ${\boldsymbol\psi}(T_\gamma/m)={\boldsymbol\psi}(0)\exp(-{\rm i}\alpha/m)$. Then ${\boldsymbol\psi}(t)$ obviously still satisfies Eq.~(\ref{eq:pseudoperidicity}). However, after the pseudo-period $T_{\rm ppo}=T_{\gamma}/m$ the primitive orbit is repeated but with a different global phase. Thus when naively integrating the global phase from $0$ to $2\pi$ and the time from $0$ to $T_\gamma$, one and the same orbit is counted $m$ times.

On the other hand, as discussed in Appendix \ref{app:uniqueness_phase}, for a given pseudo-periodic orbit, the primitive phase difference $\alpha$, {\it i.e.~}the phase difference after the primitive pseudo-period, is unique. That is that any time $T^\ast$, for which ${\boldsymbol\psi}(T^\ast)={\boldsymbol\psi}(0)\exp(-{\rm i}\alpha^\ast)$, has to satisfy $T^\ast=mT_{\rm ppo}$ with $m\in\mathbb{N}$. Thus obviously also $\alpha^\ast=m\alpha$.

Therefore,
\begin{equation}
\int{\rm d}^2q_{\parallel}\left|\frac{\partial\Re\left({\boldsymbol\psi}_{\parallel}\left(t_{\gamma}\right){\rm e}^{{\rm i}\alpha}\right)}{\partial\left(T_{\gamma},\theta\right)}\right|^{-1}=\frac{2\pi T_{\gamma}}{m}=2\pi T_{\rm ppo}.
\label{eq:parallel_integrals}
\end{equation}
The last remaining integration over $\alpha$ can straightforwardly be computed in stationary phase approximation. The stationarity condition selects those trajectories, for which the given number of particles $N_{\gamma}$ is related to the total number of particles,
\begin{equation}
N_{\gamma}=N+\frac{L}{2},
\end{equation}
however, when evaluating the integral, one should keep in mind that $\alpha=\theta-\theta(T_\gamma)$, where $\theta(T_\gamma)$ is the global phase at final time.

Finally, the oscillatory part of the many-body DOS for fixed total number of particles then reads
\begin{equation}
\tilde{\rho}_N(E)=\sum\limits_{\rm po}\frac{T_{\rm ppo}}{\pi\hbar\sqrt{\left|M_{\rm po}-1\right|}}\cos\left(\frac{1}{\hbar}S_{\rm po}(E)-\sigma_{\rm po}\frac{\pi}{2}\right).
\label{eq:trace-formula}
\end{equation}
Here the sum runs over the families of pseudo-periodic orbits satisfying
\begin{equation}
{\boldsymbol\psi}\left(T_{\rm po}\right)={\boldsymbol\psi}\left(0\right){\rm e}^{-{\rm i}\alpha_{\rm po}},
\label{eq:pseudoperiodicity-final}
\end{equation}
where $T_{{\rm po}}$ is the flying time of the orbit, which may be any integer multiple of the primitive period $T_{\rm ppo}$, defined as the smallest time, for which Eq.~(\ref{eq:pseudoperiodicity-final}) is satisfied, and $\alpha_{\rm po}$ is an arbitrary global phase depending only on the trajectory.

The argument of the cosine is given by the classical action
\begin{equation}
S_{\rm po}(E)=\hbar\int\limits_{0}^{T_{po}}{\boldsymbol\theta}\left(t\right)\cdot\dot{\bf n}\left(t\right){\rm d}t+2\pi\hbar k_{\rm po}
\label{eq:periodic_actions}
\end{equation}
and the (integer) Maslov index
\begin{equation}
\sigma_{po}=\mu_{\rm po}+\nu_{\rm po}+L\eta\left(\alpha_{\rm po}\right)-\frac{1}{2}{\rm sign}\frac{\partial N_{\gamma}}{\partial\alpha_{\rm po}},
\label{eq:Maslov_orig}
\end{equation}

We would like to give a final remark about the appearance of the global phase difference $\alpha_{\rm po}$. When considering the orbits in the reduced space, where not only the number of particles is fixed but also the global phase is set constant, they would be strictly periodic. However, as it was already remarked in \cite{traceformula_cont_sym}, a trajectory, which is periodic in reduced space may not be periodic in the full space.

On the other hand, one might have expected this behavior already in advance, since even if the nonlinear wave at final time differs from the initial one by a global phase factor, the following time evolution is again the same as the initial one.
\subsection{The Maslov index}
\label{subsec:maslov}
While Eq.~(\ref{eq:Maslov_orig}) in principle yields the correct Maslov index, it is not very helpful when calculating it in practice. A more useful formula can be obtained by not performing both integrations over ${\bf p}$ and ${\bf q}$ in Eq.~(\ref{eq:dos_before_spa}) together but one after the other. For instance, if ${\bf p}$ is integrated out first, the intermediate result for the resolvent is given by
\begin{equation}
\begin{split}
g_{\rm N}(E)=&\frac{1}{2\pi{\rm i}\hbar\left(-2\pi{\rm i}\hbar\right)^\frac{L-1}{2}}\int\limits_{0}^{2\pi}{\rm d}\alpha{\rm e}^{-{\rm i}\alpha\left(N+\frac{L}{2}\right)}\int{\rm d}^Lq \\
&\sum\limits_{\gamma}\left.\sqrt{\det\left(\begin{array}{cc} \frac{\partial^2\tilde{W}_{\gamma}}{\partial{\bf q}\partial{\bf q}^\prime} & \frac{\partial^2\tilde{W}_{\gamma}}{\partial{\bf q}\partial E} \\ \frac{\partial^2\tilde{W}_{\gamma}}{\partial E\partial{\bf q}^\prime} & \frac{\partial^2\tilde{W}_{\gamma}}{\partial E^2} \end{array}\right)}\right|_{{\bf q}^\prime={\bf q}}{\rm e}^{\frac{\rm i}{\hbar}\tilde{W}_{\gamma}},
\end{split}
\label{eq:resolvent_p_integrated}
\end{equation}
where now the trajectories satisfy the boundary conditions
\begin{equation}
\begin{split}
\Re{\boldsymbol\psi}(0)&={\bf q}\\
\Re{\boldsymbol\psi}(T_{\gamma}){\rm e}^{{\rm i}\alpha}&={\bf q},
\end{split}
\end{equation}
and their actions are given by
\begin{equation}
\tilde{W}_{\gamma}=W_{\gamma}+\frac{\hbar}{\cos\alpha}\left\{\Im{\boldsymbol \psi}(T_{\gamma})\cdot{\bf q}+\left[{\bf q}^2+\left(\Im{\boldsymbol \psi}(T_\gamma)\right)^2\right]\sin\alpha\right\}.
\end{equation}
Performing the remaining integrals in stationary phase approximation (except of those along the trajectory and along the global phase) must finally again yield Eq.~(\ref{eq:trace-formula}). However, this way following \cite{maslov_index_geometrical}, the Maslov index is given by a sum of two terms
\begin{equation}
\sigma_{\rm po}=\mu_{\rm po}^{\prime}+\nu_{\rm po}^{\prime},
\end{equation}
where $\mu_{\rm po}$ is increased and decreased by one every time the determinant of
\begin{equation}
\left(\frac{\partial\Im{\boldsymbol\psi}_{\perp}(t){\rm e}^{{\rm i}\alpha}}{\partial\Im{\boldsymbol\psi}_{\perp}(0)}\frac{\partial\Im{\boldsymbol\psi}_{\perp}(t){\rm e}^{{\rm i}\alpha}}{\partial\Re{\boldsymbol\psi}_{\perp}(0)}\right)^{-1}
\end{equation}
changes sign as a function of $t$. In fact, $\mu_{\rm po}^{\prime}$ is not an invariant property of the pseudo-periodic orbit, but depends on the choice of the initial point. $\nu_{\rm po}^\prime$ also depends on this choice and is determined by the zeros of the determinant of $\left(\frac{\partial\Im{\boldsymbol\psi}_{\perp}(t){\rm e}^{{\rm i}\alpha}}{\partial\Re{\boldsymbol\psi}_{\perp}(0)}\right)^{-1}$. This index can be determined as follows \cite{maslov_index_geometrical}: When shifting the initial point along the orbit, a caustic, which is the point at which $\left(\frac{\partial\Im{\boldsymbol\psi}_{\perp}(t){\rm e}^{{\rm i}\alpha}}{\partial\Re{\boldsymbol\psi}_{\perp}(0)}\right)^{-1}$ is zero, can appear or disappear. At such a point, $\nu_{\rm po}^\prime$ is incremented or decremented by one. This way, $\sigma_{\rm po}$ is independent of the choice of the initial point.

\subsection{Reduced vs. Extended phase space approaches}
\label{subsec:RedExt}

We conclude the presentation of the derivation of the trace formula for second-quantized many-body systems with a remark concerning the implementation of the gauge symmetry responsible of the conservation of $N$. At first glance one may think that our choice of using the periodic orbits in the extended phase-space, thus rendering them pseudo-periodic, leads to substantial technical complications compared with a construction based on periodic orbits in the reduced phase-space fixed by the total number of particles. The classical mean-field equations get, however, extremely involved when one explicitly uses the conservation of $N$ to reduce the dimensionality of the problem, as it can be easily seen from Eqns.~(\ref{eq:action_fixed_alphaJD}) and (\ref{eq:HMF}). Explicit use of $N$ as an external parameter leads then to equations of motion which are non-polynomial in the fields, thus rendering both analytical and numerical calculations much more difficult already in the non-interacting case while, with our choice, the simplicity of the mean-field equations is preserved. This fact will be more evident and crucial in the next section where we completely and {\it exactly} solve the free-field case, something which is possible because of the strict linearity of the problem when formulated in the language of pseudo-periodic orbits. 
\section{The free field}
\label{sec:FF}

The trace formula (\ref{eq:trace-formula}) for Bose-Hubbard systems finds its most natural application in the case where the mean-field equations display (discrete) field chaos going along with isolated unstable periodic solutions. Technically, this stems from the essential step where the integrations involved in the calculation of the trace are performed in stationary phase approximation. Implicitly, we are assuming that periodic orbits are isolated and do not come in continuous families. The presence of continuous families of periodic orbits is a hallmark of classical integrability \cite{Gutbook}, and therefore the trace formula cannot usually be applied to integrable systems.

In order to study the possible application of the trace formula (\ref{eq:trace-formula}) in the non-interacting limit of a discrete bosonic field we must check whether this limit corresponds to a classical integrable system or not. In view of Eqs.~(\ref{eq:QH}) and (\ref{eq:rep}), the mean field Hamiltonian corresponding to the quantum mechanical free field Hamiltonian,
\begin{equation}
\hat{H}=\sum_{ij}H_{ij}\hat{a}_{i}^{\dagger}\hat{a}_{j},
\end{equation}
is given by
\begin{equation}
H^{({\rm MF, Free})}\left({\boldsymbol \psi}^\ast,{\boldsymbol \psi}\right)=\sum_{ij}H_{ij}\left(\psi_{i}^{\ast}\psi_{j}-\frac{\delta_{ij}}{2}\right)
\end{equation}
where the term $1/2$ came from the Weyl ordering of operators implicit in our derivation of the semiclassical propagator.

First, we will show that $H^{({\rm MF, Free})}$ admits a set of $L$ independent constants of motion, implying by definition integrability. In exact analogy with the quantum case, we consider a transformation 
\begin{equation}
\psi_{i}=\sum_{\chi}u_{i\chi}\phi_{\chi}
\end{equation}
which is canonical if and only if the matrix ${\bf u}$ with entries $u_{i\chi}$ is unitary. It is a simple exercise to show that if the matrix ${\bf u}$ diagonalizes the matrix ${\bf H}$, i.e,
\begin{equation}
\sum_{ij}u_{i\chi}^{\ast}H_{ij}u_{j\chi'}=e_{\chi}\delta_{\chi\chi'}
\end{equation}
then the functions
\begin{equation}
\label{eq:Occn}
n_{\chi}\left({\boldsymbol \psi}^\ast,{\boldsymbol \psi}\right):=\sum_{ij}u_{i\chi}^{\ast}u_{j\chi}\psi_{i}^{\ast}\psi_{j}{\rm \ , \ for \ }\chi=1,\ldots,L
\end{equation}
constitute $L$ independent constants of motion under the Hamiltonian flow induced by $H^{({\rm MF, Free})}$. These classical phase-space functions are the obvious classical analogues of the quantum mechanical number operators counting excitations in the eigenstates of the single particle Hamiltonian.

Since the free mean-field Hamiltonian is integrable, the trace formula should in principle be modified to account for the continuous families of periodic orbits typical of integrable systems. Remarkably, it turns out that the non-interacting limit of a quantum field theory is not typical at all. The reason is that, as it is obvious from the quadratic dependence of $H^{({\rm MF, Free})}\left({\boldsymbol \psi}^{\ast},{\boldsymbol \psi}\right)$ on the canonical variables ${\boldsymbol \psi}^\ast$ and ${\boldsymbol \psi}$, the free field is not only integrable but it is actually {\it harmonic}. Harmonic systems are not generic integrable systems. In fact, depending on the number-theoretical relation between the energies $e_{\alpha}$ of the single particle orbitals, they share some fundamental properties of the chaotic case. In particular, if the single-particle energies are not commensurable (the generic situation for a randomly chosen matrix ${\bf H}$), the periodic orbits of the system are actually isolated. To understand this we focus on the solutions of the classical limit, which is just the single-particle, linear Schr\"odinger equation
\begin{equation}
\label{eq:SPSE}
i\hbar \frac{d}{dt}\psi_{i}(t)=\sum_{j}H_{ij}\psi_{j}(t),
\end{equation}
with solution
\begin{equation}
\label{eq:SPSES}
{\boldsymbol \psi}(t)={\rm e}^{-\frac{{\rm i}}{\hbar}{\bf H}t}{\boldsymbol \psi}(0).
\end{equation}
Note that the eigenvector ${\bf v}^{(\chi)}$ of ${\bf H}$ 
\begin{equation}
{\bf H}{\bf v}^{(\chi)}=e_{\chi}{\bf v}^{(\chi)}
\end{equation}
with eigenvalue 
\begin{equation}
e_{\chi}=\hbar w_{\chi}
\end{equation}
defines a family of periodic orbits with fundamental frequency $w_{\chi}$
\begin{equation}
\label{eq:PO1}
{\bf v}^{(\chi)}(t)={\rm e}^{-{\rm i}w_{\chi}t}{\bf v}^{(\chi)}.
\end{equation}
To show that these are the only periodic orbits of the system and that they are indeed isolated we note that, because of linearity, Eq.~(\ref{eq:SPSES}) can be expressed as a linear combination,
\begin{equation}
{\boldsymbol \psi}(t)=\sum_{\chi}c_{\chi}(\psi(0)){\rm e}^{-{\rm i}w_{\chi}t}{\bf v}^{(\chi)}
\end{equation}
for some constants $c_{\chi}(\psi(0))$ depending only on the initial condition ${\boldsymbol \psi}(0)$. Assume now that for this initial condition there is a pseudo-periodic solution with period $T$, namely, that 
\begin{equation}
{\boldsymbol \psi}(T)={\boldsymbol \psi}(0)\exp(-{\rm i}\alpha).
\end{equation}
Comparing the eigenvector expansions of both sides of this equation we get the consistency condition
\begin{equation}
c_{\chi}(\psi){\rm e}^{-{\rm i}\alpha}=c_{\chi}(\psi){\rm e}^{-{\rm i}w_{\chi}T} {\rm \ for \ all \ }\chi
\end{equation}
which for incommensurable frequencies $w_{\chi}$ can be only satisfied if $T$ satisfies
\begin{equation}
\label{eq:PeriodJD}
T=T_{\tilde{\chi}}:=\frac{\alpha}{w_{\tilde{\chi}}}
\end{equation}
for some $\tilde{\chi}$ and simultaneously $c_{\chi}(\psi)=\delta_{\chi\tilde{\chi}}$ \footnote{In fact, for certain values of $\alpha$, certain combinations of integers $k$ and $k^\prime$ exist such that for a pair of families $\chi,\chi^{\prime}$ the condition $\alpha=2\pi\frac{w_{\chi^\prime}k^{\prime}-w_{\chi}k}{w_{\chi}-w_{\chi^{\prime}}}$ is satisfied. Then, the pseudo-periodic orbits are no longer isolated. However, these values of $\alpha$ are discrete and therefore have zero measure in the final integration.}. This means that for a generic matrix ${\bf H}$, the {\it only} periodic orbits are the ones emerging from the eigenstates of the single-particle problem. For fixed energy they are obviously discrete, and therefore isolated. 

It is important to stress that, being simply the classical limit of the theory, there is no physical reason whatsoever to prefer a {\it normalized} solution of the equations of motion. In fact, each eigenvector ${\bf v}^{(\chi)}$ defines a complete, continuous family of periodic orbits with norms that vary continuously. As it will be clear below, this continuous family corresponds to the expected continuous dependence of the action with the energy $E$. For fixed $E$ in the trace formula, a specific value of the norm $\left|{\bf v}^{(\chi)}\right|^{2}$ will be selected. 

Note here that in this special case of zero interactions the integration along the periodic orbit is actually the same as the one along the initial global phase, due to Eq.~(\ref{eq:PO1}), namely the time evolution for a periodic orbit is simply a change in the global phase. Thus, contrary to Sec.~\ref{subsec:osc}, only one integration, namely the one along the trajectory, has to be performed exactly. The evaluation of the integrals in Eq.~(\ref{eq:dos_before_spa}) is then strictly equivalent to the standard derivation of the trace formula \cite{Gutzwiller_trace} and yields for the oscillatory part of the many-body DOS for non-interacting systems
\begin{eqnarray}
\label{eq:dosfreefield}
\tilde{\rho}_N(E)&=&-\frac{1}{2\pi}\Im\int\limits_{0}^{2\pi}d\alpha{\rm e}^{-{\rm i}\alpha \left(N+\frac{L}{2}\right)} \\ &\times& \sum\limits_{\rm po}\frac{T_{\rm ppo}}{{\rm i}\pi\hbar\sqrt{\left|M_{\rm po}-1\right|}}\exp\left(\frac{\rm i}{\hbar}\tilde{S}_{\rm po}-{\rm i}\sigma_{\rm po}\frac{\pi}{2}\right). \nonumber
\end{eqnarray}
Here, the pseudo-periodic orbits and their actions are still given by Eqns.~(\ref{eq:pseudoperidicity}) and (\ref{eq:action_fixed_alpha}), while the stability matrix $M_{\rm po}$ is given by Eq.~(\ref{eq:stability-matrix}) but with the perpendicular coordinates increased by one further dimension. Also, due to the coincidence of the integration along the orbit and along the global phase, the primitive period $T_{\rm ppo}$ is now determined by a \emph{full} cycle, {\it i.e.}~the smallest (non-zero) time for which ${\boldsymbol\psi(t)}={\boldsymbol\psi(0)}$. The Maslov index $\sigma$ can still be calculated according to Sec.~\ref{subsec:maslov}.

Since the periodic orbits are isolated, our trace formula can be applied directly. In the non-interacting case it is instructive to perform the integration over $\alpha$ (responsible for selecting orbits with fixed given total number of particles $N$) exactly. The pseudo-periodic orbits are organized in $L$ families corresponding to the $L$ different eigenvectors of the matrix ${\bf H}$. Consider first the primitive pseudo-periodic orbit associated with the eigenvector $\chi$, whose time-dependence can be explicitly constructed as in Eq.~(\ref{eq:PO1}) and therefore has frequency $w_{\chi}$. As characteristic of harmonic systems, the period is independent of the energy $E$ of the trajectory, which is given by
\begin{eqnarray}
\label{eq:EJD}
E&=&\sum_{ij}\left(v^{(\chi)}_{i}\right)^{\ast}H_{ij}v^{(\chi)}_{j}-\frac{1}{2}{\rm Tr~}{\bf H} \nonumber \\ &=&e_{\chi}\left|{\bf v}^{(\chi)}\right|^{2}-\frac{1}{2}\sum_{\chi}e_{\chi}.
\end{eqnarray}
Note that the energy $E$ appearing in the trace formula and as the argument of the many-body DOS has nothing to do a priori with the eigenvalues $e_{\chi}$ of the single-particle problem, beyond the fact that the spectrum of ${\bf H}$ is part of the parameters that define the many-body problem. From the point of view of semiclassics in second-quantized systems, in the non-interacting case the single particle energies $e_{\chi}$ simply provide the frequencies $w_{\chi}$ of the harmonic problem that defines the classical limit.
  
Using Eqs.~(\ref{eq:action_fixed_alpha}) and (\ref{eq:action_fixed_alphaJD}) the action of the $k$th repetition of any member of the $\chi$-family is easily found to be
\begin{equation}
\tilde{S}_{\chi}^{(k)}=(\alpha+2\pi k)\hbar\left|{\bf v}^{(\chi)}\right|^{2},
\end{equation}
and therefore, using Eq.~(\ref{eq:EJD}) we get the action 
\begin{equation}
\tilde{S}_{\chi}^{(k)}(E)=\frac{\alpha+2\pi k}{w_{\chi}}\left(E+\frac{\hbar}{2}\sum_{\chi'}w_{\chi'}\right)
\end{equation}
and the period
\begin{equation}
T_{\chi}^{(k)}(E)=\frac{\alpha+2\pi k}{w_{\chi}}
\end{equation}
of the pseudo periodic orbits in terms of the energy. Following the discussion below Eq.~(\ref{eq:dosfreefield}), the period of the primitive pseudo orbits is then given simply by $2\pi/w_{\chi}$.

The next step is the calculation of the stability matrices and Maslov indexes entering the trace formula Eq.~(\ref{eq:dosfreefield}). This is a standard exercise for harmonic systems, and we present it here only to illustrate the conceptual relation between eigenstates of single particle problems in first quantization and the semiclassical approach to second-quantized many-body systems of indistinguishable particles.

The stability matrix is given by the local properties of the classical evolution around a specific pseudo periodic orbit, Eq.~(\ref{eq:SPSES}), as the linear transformation relating small initial ${\boldsymbol \delta v}^{(\chi)}(0)$ and final ${\boldsymbol \delta v}^{(\chi)}(T_{\chi}^{(k)})$ deviations from the reference orbit. Using the linearity of the classical dynamics we easily get 
\begin{eqnarray}
{\boldsymbol \delta v}^{(\chi)}(T_{\chi}^{(k)})&=&{\rm e}^{-\frac{{\rm i}}{\hbar}{\bf H}T_{\chi}^{(k)}}{\boldsymbol \delta v}^{(\chi)}(0)  \\
&=&\sum_{\chi'}{\bf v}^{(\chi')} \cdot {\boldsymbol \delta v}^{(\chi)}(0){\rm e}^{-{\rm i}(\alpha+2\pi k)\frac{e_{\chi'}}{e_{\chi}}} {\bf v}^{(\chi')}. \nonumber
\end{eqnarray}
In view of Eq.~(\ref{eq:stability-matrix}), we then obtain for the components of the deviations along the directions $\chi'=\chi^{\perp}$ perpendicular to the $\chi$-orbit 
\begin{equation}
\delta v^{(\chi)}_{\chi'}(T_{\chi}^{(k)}){\rm e}^{i\alpha}={\rm e}^{-{\rm i}(\alpha+2\pi k)\frac{e_{\chi'}}{e_{\chi}}+{\rm i}\alpha}\delta v^{(\chi)}_{\chi'}(0).
\end{equation}   
This equation indicates that the stability matrix is obviously block-diagonal, with the block corresponding to $\chi'$ being simply a rotation matrix with angle 
\begin{equation}
\theta^{\chi,(k)}_{\chi'}=(\alpha+2\pi k) \frac{e_{\chi'}}{e_{\chi}}-\alpha,
\end{equation} 
and therefore
\begin{equation}
\left|M_{\chi}^{(k)}-1\right|=\prod_{\chi' \ne \chi}\left|\det\left(
\begin{array}{cc}
\cos\theta^{\chi,(k)}_{\chi'}-1 & \sin\theta^{\chi,(k)}_{\chi'} \\
-\sin\theta^{\chi,(k)}_{\chi'} & \cos\theta^{\chi,(k)}_{\chi'}-1
\end{array}
\right)\right|.
\end{equation} 
This yiealds eventually
\begin{equation}
\sqrt{\left|M_{\chi}^{(k)}-1\right|}=\prod_{\chi' \ne \chi}2\left|\sin \left(\frac{\alpha+2\pi k}{2}\frac{e_{\chi'}}{e_{\chi}}-\frac{\alpha}{2}\right)\right|.
\end{equation}
Moreover, the Maslov index for the $k$-th repetition of the $\chi$-family is given by \cite{traceformula_ho}
\begin{equation}
\sigma_{\chi}^{(k)}=2k+2\sum\limits_{\chi^\prime\neq\chi}\left\lfloor\left(k+\frac{\alpha}{2\pi}\right)\frac{e_{\chi^\prime}}{e_{\chi}}-\frac{\alpha}{2\pi}\right\rfloor+1,
\end{equation}
where $\lfloor x\rfloor$ denotes the integer part of $x$.

Substitution of the actions, stabilities and Maslov indexes in Eq.~(\ref{eq:dosfreefield}) leads to the semiclassical trace formula for the free bosonic field with fixed total number of particles $N$ as a sum over pseudo-periodic orbits and their repetitions:
\begin{eqnarray}
\label{eq:FinJDI}
\tilde{\rho}_N(E)&=&-\frac{\Re}{2\pi}\int\limits_{0}^{2\pi}d\alpha{\rm e}^{-{\rm i}\alpha N} \sum_{\chi=1}^{L}\frac{1}{e_{\chi}}\\ &\times& \sum_{k=1}^{\infty}\frac{{\rm e}^{{\rm i}\left[\frac{\alpha+2\pi k}{e_{\chi}}\left(E+\frac{1}{2}\sum_{\chi'}e_{\chi'}\right)-\sigma_{\chi}^{(k)}\frac{\pi}{2}\right]}}{\prod_{\chi' \ne \chi}2\left|\sin \left(\frac{\alpha+2\pi k}{2}\frac{e_{\chi'}}{e_{\chi}}-\frac{\alpha}{2}\right)\right|}. \nonumber
\end{eqnarray}
If necessary, the last integration over $\alpha$ can be performed by taking $\alpha=0$ in all smooth terms, and calculating exactly the integral involving the highly oscillatory ones, and this may be indeed the way to proceed for specific calculations based on the pseudo-periodic orbits.

For the smooth (Weyl) contribution we left out again the $\alpha$-integration and easily get
\begin{eqnarray}
\label{eq:WeylJD}
\bar{\rho}_N(E)&&=\frac{\Re}{2\pi}\int\limits_{0}^{2\pi}d\alpha{\rm e}^{-{\rm i}\alpha N} \\ 
\left(\frac{4}{\pi}\right)^L&&\int{\rm d}^Lq\int{\rm d}^Lp\delta\left[E-H^{(MF)}({\bf p},{\bf q})-\alpha({\bf q}^2+{\bf p}^2)\right]. \nonumber
\end{eqnarray}

For completeness, we show below the consistency of the trace formula, Eq.~(\ref{eq:FinJDI}), with the semiclassical quantization procedure for direct quantization of invariant manifolds in phase space, so-called EBK-quantization, valid only for integrable systems. In the following we apply it to the  case of interest here.  

Within EBK quantization, the classical Hamiltonian is first written in terms of a new set of canonical variables $({\boldsymbol \phi},{\bf J})$ where ${\bf J}={\bf J}({\bf n})$ are combinations of the classical constants of motion, such that $H({\boldsymbol \phi},{\bf J})=H({\bf J})$. In our case, these functions are simply given by ${\bf J}=\hbar{\bf n}({\boldsymbol \psi})$ with $n_{\chi}({\boldsymbol \psi})$ defined in Eq.~(\ref{eq:Occn}). In the new variables the Hamiltonian is given by
\begin{equation}
H({\bf J})=\sum_{\chi}\left(J_{\chi}-\frac{\hbar}{2}\right)w_{\chi}.
\end{equation}  
In a second step, EBK quantization selects the values of the classical actions ${\bf J}$ such that
\begin{equation}
J_{\chi}=\hbar\left(n_{\chi}+\frac{\beta_{\chi}}{4}\right) 
\end{equation}
with $n_{\chi}=0,1,2,\ldots$ and with indexes $\beta_{\chi}$ given by the number of turning points of any classical trajectory evolving in the phase space manifold defined by the set of quantized constants of motion, so in our case $\beta_{\chi}=2$ for all $\chi$. The EBK-quantized energies are then obtained by
\begin{equation}
E_{\bf n}=H\left[{\bf J}=\hbar\left(n_{\chi}+\frac{1}{2}\right)\right],
\end{equation}
giving for our case
\begin{equation}
E_{\bf n}=\sum_{\chi}n_{\chi}e_{\chi},
\end{equation}
providing a proper and physical interpretation of the EBK quantization condition in the context of the free bosonic field: quantization of the many-body energy levels is due to quantization of the occupation numbers. As it is well known \cite{Brackbook}, the EBK quantization of linear (harmonic) systems is exact, and indeed this is the exact quantum mechanical spectrum of this system.

So far, we used a well known quantization method to derive a well known result in the framework of first-quantized systems. What makes the second-quantized approach rather special is the status of the phase-space observable representing the total number of particles $N$. The key point is that the conserved quantity
\begin{equation}
N({\boldsymbol \psi})=\sum_{i}|\psi_{i}(t)|^{2}
\end{equation}
plays a distinctive role in the field theoretic scenario, namely, it labels subspaces of given total number of particles. Note that such a condition is never encountered in the description of particle (instead of field) systems. There, the function  $N({\boldsymbol \psi})$ is simply the sum of actions, and there does not exist a physical interpretation as an observable. 

This detail makes the semiclassical approach for fields with conserved number of particles conceptually different from its interpretation as a set of first-quantized harmonic oscillators. In particular, if one wants to study the many-body spectrum with a given, fixed $N$, one must project the EBK DOS. This is again accomplished by introducing a variable $\alpha$, playing the role of an imaginary chemical potential,
\begin{eqnarray}
\label{eq:FinJDII}
\rho_N(E)&=&\frac{1}{2\pi}\int\limits_{0}^{2\pi}d\alpha{\rm e}^{-{\rm i}\alpha N} \\ &\times& \sum_{{\bf n}}\delta\left(E-\sum_{\chi}n_{\chi}e_{\chi}\right){\rm e}^{{\rm i}\alpha\sum_{\chi}n_{\chi}}. \nonumber
\end{eqnarray}
In order to transform this sum over quantum numbers corresponding to quantized occupations into a trace formula where periodic (or pseudo-periodic) orbits appear, we proceed in a similar way as in the semiclassical quantization of harmonic systems and introduce
\begin{eqnarray}
{\cal Z}(\beta,\alpha)&=&\int_{0}^{\infty}dE{\rm e}^{-\beta E}\sum_{{\bf n}}\delta\left(E-\sum_{\chi}n_{\chi}e_{\chi}\right){\rm e}^{{\rm i}\alpha\sum_{\chi}n_{\chi}} \nonumber \\
&=&\sum_{{\bf n}}{\rm e}^{-\beta\sum_{\chi}n_{\chi}e_{\chi}}{\rm e}^{{\rm i}\alpha\sum_{\chi}n_{\chi}}.
\end{eqnarray}
Performing the sums given as geometric series, we get
\begin{equation}
{\cal Z}(\beta,\alpha)=\prod_{\chi'}\left[1-{\rm e}^{-(\beta e_{\chi'}-{\rm i}\alpha)}\right]^{-1}
\end{equation}
which is suitable to compute the inverse Laplace transform required in Eq.~(\ref{eq:FinJDII})
\begin{eqnarray}
\label{eq:ZJD}
&&\sum_{{\bf n}}\delta\left(E-\sum_{\chi}n_{\chi}e_{\chi}\right){\rm e}^{{\rm i}\alpha\sum_{\chi}n_{\chi}} \\&&=\frac{1}{2\pi{\rm i}}\int_{\epsilon-{\rm i}\infty}^{\epsilon-{\rm i} \infty}d\beta{\rm e}^{\beta E}{\cal Z}(\beta,\alpha), \nonumber
\end{eqnarray}
by means of the Bromwich formula where $\epsilon$ is real and positive and chosen such that it is larger than the real part of all the poles of ${\cal Z}(\beta,\alpha)$. These poles are easily found to be located at
\begin{equation}
\label{eq:PolesJD}
\beta_{\chi}^{(k)}={\rm i}\left(\frac{\alpha +2\pi k}{e_{\chi}}\right){\rm \ \ , \ \ }k=0,\pm 1,\pm 2,\ldots
\end{equation}
and are naturally labeled by the index $\chi$ of the single-particle energy that, as we know, also denotes periodic orbits. The second index $k$ labeling the positions of the poles is naturally interpreted within the trace formula as the repetition of the pseudo-periodic orbits as well. 

Having at hand our trace formula, Eq.~(\ref{eq:FinJDI}), we are ready to identify and give physical interpretation to the different factors appearing in the ``exact" trace formula obtained by evaluating the contributions from the residua of Eq.~(\ref{eq:ZJD}) at the poles given by Eq.~(\ref{eq:PolesJD}). Consider first
\begin{equation}
\left. {\rm e}^{\beta E}\right|_{\beta=\beta_{\chi}^{(k)}}={\rm e}^{{\rm i}\frac{\alpha +2\pi k}{e_{\chi}}E}
\end{equation}
giving the energy-dependent term in the action of Eq.~(\ref{eq:FinJDI}). Now  
\begin{eqnarray}
\left. \left(1-{\rm e}^{-( \beta e_{\chi'}-{\rm i}\alpha)}\right)^{-1}\right|_{\beta=\beta_{\chi}^{(k)}} &=&\left[1-{\rm e}^{-{\rm i}\left((\alpha+2\pi k)\frac{e_{\chi'}}{e_{\chi}}-\alpha\right)}\right]^{-1} \nonumber \\
&=&\frac{{\rm e}^{{\rm i}\left(\frac{\alpha+2\pi k}{2}\frac{e_{\chi'}}{e_{\chi}}-\frac{\alpha}{2}\right)}}{2{\rm i}\sin\left(\frac{\alpha+2\pi k}{2}\frac{e_{\chi'}}{e_{\chi}}-\frac{\alpha}{2}\right)} 
\end{eqnarray}
give the stability factors in Eq.~(\ref{eq:FinJDI}). Putting all together, we finally get
\begin{eqnarray}
\label{eq:ZJD2}
&&\sum_{{\bf n}}\delta\left(E-\sum_{\chi} n_{\chi}e_{\chi}\right){\rm e}^{{\rm i}\alpha\sum_{\chi}n_{\chi}} \\&&=\sum_{\chi}\frac{1}{e_{\chi}}\sum_{k=-\infty}^{\infty} \frac{{\rm e}^{{\rm i}\left[\frac{\alpha+2\pi k}{e_{\chi}}\left(E+\frac{1}{2}\sum_{\chi'}e_{\chi'}\right)-\frac{L\alpha}{2}-k\pi\right]}}{\prod_{\chi' \ne \chi}2{\rm i}\sin \left(\frac{\alpha+2\pi k}{2}\frac{e_{\chi'}}{e_{\chi}}-\frac{\alpha}{2}\right)}, \nonumber
\end{eqnarray}
in full agreement with our trace formula, Eq.~(\ref{eq:FinJDI}). Note, however, that the terms with $k=0$ must be computed independently, and in the limit $\alpha=0$, appropriate for the asymptotic regime $N\to \infty$, they precisely provide the Weyl term, Eq.~(\ref{eq:WeylJD}). 

\section{Extensions and relation with previous approaches}
\label{sec:EXT}

The semiclassical trace formula provides a fundamental and rigorous connection between the spectrum of a quantum system and the features and properties of its classical limit. As such, it should be able to cover less general approximations that rely on classical information to explain features of the quantum mechanical DOS. Note, however, that the trace formula does not associate directly dynamical properties of the classical system with individual energy levels of  the quantum spectrum. The correct association is between periodic solutions of the classical mean field equations and Fourier components of the full many-body quantum DOS as a function of the energy. Semiclassically, the emergence of discrete energies is an interference phenomenon due to the coherent superposition of all these harmonics. 

In this paper, we have addressed the situation where the classical mean field dynamics, understood as a dynamical system, is such that all the periodic solutions are isolated. Remarkably, while this is usually the case that only for fully chaotic dynamics, the important case of a non-interacting bosonic field falls into this category as well. This is because the periodic solutions of the corresponding classical limit, which is harmonic, are again isolated. 

As it is well known \cite{Gutbook,Tabor} a generic dynamical system is actually neither integrable nor chaotic, and the trace formula and more generally, semiclassical quantization methods for first-quantized systems were correspondingly generalized in order to describe also the integrable-to-chaotic transition \cite{OdA}. Also, extensions and generalizations of the trace formula can be used to quantize selected, specific locally integrable dynamics of the classical phase space. In the following we present a brief discussion of the connections between the approach presented here and other semiclassical methods aiming to associate features of the many-body quantum spectrum with special classical structures. 

The first, non-generic, situation appears if the classical mean-field equations admit a static solution ${\boldsymbol \psi}^{({\rm fix})}$ defined by

\begin{eqnarray}
\left.{\rm i}\hbar\dot{\boldsymbol \psi}(t)\right|_{{\boldsymbol \psi}={\boldsymbol \psi}^{({\rm fix})}}&=&\left.\frac{\partial H^{({\rm MF})}\left({\boldsymbol \psi}^\ast(t),{\boldsymbol \psi}(t)\right)}{\partial{\boldsymbol \psi}^\ast(t)}\right|_{{\boldsymbol \psi}={\boldsymbol \psi}^{({\rm fix})}} \nonumber \\ &=& 0
\label{eq:FP}
\end{eqnarray}
such that the classical mean-field motion is strictly stable in all directions around ${\boldsymbol \psi}^{({\rm fix})}$. In this case, the spectrum of the system contains energy levels associated with the quantization of the normal modes describing small oscillations around the stable fixed point. Long-wavelength oscillations of the DOS due to this sequence of energy levels will be described by our trace formula for the harmonic system obtained by the quadratic expansion of  $H^{({\rm MF})}\left({\boldsymbol \psi}^\ast(t),{\boldsymbol \psi}(t)\right)$ around ${\boldsymbol \psi}(t)={\boldsymbol \psi}^{({\rm fix})}$. In the context of Bose-Hubbard models, this result corresponds to the well known Bogoliubov approximation \cite{Bogoliubov-approximation}, and by its very construction it is valid only for energies $E$ such that  $H^{({\rm MF})}\left({\boldsymbol \psi}^{{\rm fix,*}},{\boldsymbol \psi}^{{\rm fix}}\right) \simeq E$.

Generic dynamical systems display a mixed phase space with a mixture of regular (locally integrable) and chaotic motion, often with complicated fractal structures along their borders \cite{Tabor}. None of the versions of the trace formula presented here are strictly valid for this situation. However, the whole machinery developed to incorporate bifurcation effects characteristic of the mixed dynamics scenario in the trace formulas for first-quantized systems \cite{Brackbook} can be directly imported into the many-body context. The semiclassical quantization of regular islands in phase space can be performed equivalently by constructing local constants of motion and applying the EBK quantization locally \cite{OdA}. These approaches may be very important in the field theoretical context, as numerical investigations consistently indicate that the phase space of the dynamics induced by the mean field Hamiltonian has a very complicated structure, where hard chaos might be very difficult to observe \cite{Sandro,CBH3}. 

A special situation arises for the case $L=2$ where interactions are present, rendering the classical dynamics nonlinear, but the system is still integrable . Here exist two constants of motion (the total energy and number of particles), hence as many as degrees of freedom. This two-site Bose Hubbard model has been extensively studied both quantum mechanically and classically, and can be exactly mapped into the Josephson Hamiltonian describing bosonic excitations in superconducting devices \cite{Oberthaler2006}. Semiclassically, this system has been extensively studied by means of the EBK quantization method appropriate for classically integrable systems. The construction of the classical actions and the quantization conditions is, however, substantially more complicated \cite{GraefeKorsch2007,Walter,Oberthaler}.  In order to apply the methods presented here to this situation, it is convenient to work directly in the reduced phase space obtained by fixing the total number of particles. The dynamics is now essentially one-dimensional, and our trace formula can be used, as energy-conserving one-dimensional systems possess both integrable dynamics and isolated periodic orbits. The result of this calculation is just the WKB approximation to the energy levels, which can be improved in several ways \cite{Walter,Shmuel}. This equivalence between trace formulas, EBK quantization and WKB methods for conservative one-dimensional systems is well known in the context of first-quantized semiclassics \cite{Brackbook}.  

Furthermore, our trace formula allows one to calculate the contribution to the DOS from periodic orbits with both stable and unstable local classical dynamics, a generic case for multi-dimensional systems \cite{helium}. Note that for the non-interacting limit of the Bose-Hubbard model, we encountered already the situation where the local classical flow around the {\it isolated} periodic orbits is stable. The difference between the character of the different degrees of freedom is fully encoded in the properties of the stability matrices around the classical periodic orbits, and therefore the application of the trace formula requires their explicit calculation. The spirit of this approach is not the complete enumeration of all periodic orbits, but the study of the contribution of particular solutions to the DOS. The study of such effects has been successfully carried out in the first-quantized approach to the helium atom \cite{klaus} and to the semiclassical quantization of solitons in the context of continuous quantum field theory \cite{solitons}. 

\section{Conclusions}
\label{sec:CONC}

We have presented a rigorous derivation of the semiclassical approximation for the quantum mechanical DOS of many-body quantum systems described by bosonic quantum fields on finite lattices, starting from the exact path integral form of the many-body propagator. We showed explicitly how to derive both the smooth (Weyl) and oscillatory (Gutzwiller) contributions to the DOS, and provide a trace formula for the later. Our approach follows and generalizes the original pioneering methods introduced by Gutzwiller for chaotic single-particle systems. We avoided the coherent state representation, with its characteristic need to complexify the classical limit of the theory, by using quadrature states of the field. As a special feature of the field scenario, the classical limit is a discrete classical field and its isolated (mean-field) periodic solutions are the input of the trace formula. 

Another special property is the existence of a continuous symmetry related to the conservation of the total number of particles in closed systems. We applied the methods of symmetry-projected semiclassical densities of states to get an expression for the many-body DOS within each sector with fixed total number of particles. Interestingly, due to the fact that the quantum problem in the non-interacting case reduces to a harmonic system, our trace formula is applicable since the periodic orbits can be shown to be isolated in the generic case where the single-particle energies are not commensurable, as in the chaotic case.

As for the single-particle case, our trace formula shows how the existence of discrete many-body energy levels emerging from the continuous, smooth background given by the Weyl law, is an interference phenomenon. The correct density of states and its characteristic profile made up from Dirac-delta peaks is built by the coherent effect of oscillatory contributions, one for each periodic orbit, i.e, periodic mean-field solutions.  

The study of the quantum manifestations of classical solitons (particular solutions of the non-linear equations) is matter of recent interest in the cold-atoms community. We expect that the application of our methods may help to understand and quantify interference between different solitonic contributions to the many-body DOS in terms of harmonics building up the trace formula, a pure quantum effect not to be confused with the possibility of having remnants of wave interference at the mean-field level. Specifically, if the classical non-linear field equations admit two solitonic solutions with actions $S_{1}$ and $S_{2}$ which are stable or only mildly unstable, we can approximate (up to higher repetitions) Eq.~(\ref{eq:trace-formula}) as 
\begin{equation}
\tilde{\rho}_N(E)\simeq A_{1}\cos\left(\frac{1}{\hbar}S_{1}(E)\right)+A_{2}\cos\left(\frac{1}{\hbar}S_{2}(E)\right)
\end{equation}  
and the DOS will display a characteristic beating patterns due to the interference between the two oscillatory functions, reminiscent of (super-) shell effects in the nuclear physics context \cite{shelleffects,Brackbook}. We see that this kind of analysis will provide a deeper, semiclassical understanding of long-range structures in the energy spectrum of many-body systems.   

Our work paves the way to a systematic study of the role of classical field solutions in the many-body DOS for discrete quantum fields. The close formal analogy of the many-body trace formula (\ref{eq:trace-formula}) with the Gutzwiller trace formula implies, once the conservation of total number of particles is taken care of, that the methods routinely used in the context of single-particle semiclassics can be straightforwardly generalized to show that the spectral fluctuations have universal correlations on the scale of the mean level spacing if the classical field equations are chaotic. The universal spectral correlations are linked to interference between periodic orbits with quasi-degenerate actions and periods beyond the Ehrenfest time, given by $\sim \lambda^{-1}\log \hbar^{-1}$ for single-particle systems with a classical limit with Lyapunov exponent $\lambda$. In the many-body case, correspondingly, quantum interference is additionally governed by another log-time scale, the Eherenfest time $\Lambda^{-1} \log N$. Here, $\Lambda$ denotes the average Lyapunov exponent of the assumed non-linear mean-field dynamics of the classical field limit. Many-body interference evolves at time scales beyond this Ehrenfest time and therefore is not accounted for in usual so-called Truncated Wigner approaches\footnote{see \cite{TW} for a derivation of the Truncated Wigner method along the lines of the present work}.  

Interfering quasi-degenerate periodic mean-field solutions are expected to lead to the emergence of universal many-body spectral fluctuations. While the close formal similarity to the chaotic single-particle case directly implies corresponding RMT-type expressions for the spectral many-body correlator or form factor, interesting new parametric correlation functions, involving (changes in) particle number an interaction strength, appear for the many-body case.  

The non-interacting case has, however, non-generic spectral fluctuations that do not correspond to the expected Poissonian spectra of integrable systems, a peculiar consequence of the field theoretical scenario where the free field is actually a linear, not only integrable, system. Remarkably, although for non-interacting systems the linearity of the classical limit makes it highly non-generic from the point of view of dynamical systems, this is the generic case for free fields from the field-theoretical side. Exploring the application of semiclassical techniques for the quasi-integrable case where a small interaction is treated within semiclassical perturbation theory demands then techniques adequate to this non-generic situation \footnote{J.~D.~Urbina {\it et al}, work in progress}.  

With our work we hope to contribute to the qualitative and quantitative analysis of both universal and system-specific features in the energy spectra of many-body systems by means of periodic orbit theory.

{\bf Acknowledgments}: We kindly acknowledge P. Schlagheck for fruitful discussions and his support of this project.

{\bf Note}: A derivation of a related trace formula has been performed independently by R.~Dubertrand and S.~M\"uller \cite{SebRem} using a periodic orbit theory approach in the reduced phase space, see Sec.~\ref{sec:EXT}.   

\begin{appendix}
\section{Derivation of the semiclassical prefactor of the trace formula}
\label{app:trace-prefactor}
%
The simplification of the product of the two determinants appearing in Eq.~(\ref{eq:dos_directly_after_perpendicular_integrations}) to the form in which they appear in Eq.~(\ref{eq:dos_before_parallel_integrations}) can, to a large extent, be performed by following Ref.~\cite{traceformula_cont_sym} (see Eqs.~(2.14-3.12) there). Therefore, here only the main steps will be presented, in order to show how to correctly account for the phase difference $\alpha$. For notational simplicity, in the following we will use $t$ instead of $T_{\gamma}$ to represent the period of the pseudo-periodic orbit $\gamma$.

First, guided by the Gutzwiller trace formula, where the semiclassical prefactor is determined by the monodromy matrix, we try to bring the second matrix in Eq.~(\ref{eq:dos_directly_after_perpendicular_integrations}) into a form, where ${\boldsymbol\psi}$ appears in such a form that without derivatives it could be replaced by ${\bf p}$ and ${\bf q}$. This is achieved by
\begin{equation}
\begin{split}
&\det\left(\begin{array}{cc}
\frac{\partial\Im{\boldsymbol\psi}_{\perp}(0)}{\partial{\bf q}_{\perp}}+\frac{\sin\alpha}{\cos\alpha} & \frac{\partial\Im{\boldsymbol\psi}_{\perp}(0)}{\partial\bf p}-\frac{1}{\cos\alpha} \\
\frac{\partial\Re{\boldsymbol\psi}(t)}{\partial{\bf q}_{\perp}}-\frac{1}{\cos\alpha} & \frac{\partial\Re{\boldsymbol\psi}(t)}{\partial{\bf p}}+\frac{\sin\alpha}{\cos\alpha}
\end{array}\right)= \\
&\det\frac{1}{\cos\alpha}\frac{\partial\left(\Im{\boldsymbol\psi}_{\perp}(0){\rm e}^{{\rm i}\alpha}-{\bf p}_{\perp},\Re{\boldsymbol\psi}(t){\rm e}^{-{\rm i}\alpha}-{\bf q}\right)}{\partial\left({\bf q}_{\perp},{\bf p}\right)}.
\end{split}
\end{equation}
With this the steps in Ref.~\cite{traceformula_cont_sym} can be carried over one to one in order to obtain (compare with Eq.~(3.4) therein)
\begin{equation}
\begin{split}
&\frac{\det\left(\begin{array}{cc}
\frac{\partial\Im{\boldsymbol\psi}_{\perp}(0)}{\partial{\bf q}_{\perp}}+\frac{\sin\alpha}{\cos\alpha} & \frac{\partial\Im{\boldsymbol\psi}_{\perp}(0)}{\partial\bf p}-\frac{1}{\cos\alpha} \\
\frac{\partial\Re{\boldsymbol\psi}(t)}{\partial{\bf q}_{\perp}}-\frac{1}{\cos\alpha} & \frac{\partial\Re{\boldsymbol\psi}(t)}{\partial{\bf p}}+\frac{\sin\alpha}{\cos\alpha}
\end{array}\right)}{\det\frac{\partial\left(\Im{\boldsymbol\psi}(0),t\right)}{\partial\left({\bf p},E\right)}}= \\
&\:\left(\cos\alpha\right)^{-(2L-2)}\det\left(\frac{\partial\left(\Re{\boldsymbol\psi}_{\parallel}(t){\rm e}^{-{\rm i}\alpha},E,N_{\gamma}\right)}{\partial\left(\Im{\boldsymbol\psi}_{\parallel}\left(0\right),t,\theta\right)}\right) \\
&\;\times\det\left(\frac{\partial\left(\Im{\boldsymbol\psi}_{\perp}(0){\rm e}^{{\rm i}\alpha}-{\bf p}_{\perp},\Re{\boldsymbol\psi}_{\perp}(t){\rm e}^{-{\rm i}\alpha}-{\bf q}_{\perp},\theta\right)}{\partial\left({\bf q}_{\perp},\Im{\boldsymbol\psi}_{\perp}\left(0\right),N_{\gamma}\right)}\right),
\end{split}
\end{equation}
where $\theta$ is the initial global phase of the nonlinear wave and $N_{\gamma}$ is the number of particles determined by the nonlinear wave.

Next, in the first of these two matrices, we again replace the derivative with respect to $\Im{\boldsymbol\psi}_{\perp}(0)$ by one with respect to $\Im\left({\boldsymbol\psi}_{\perp}(0)\exp\left({\rm i}\alpha\right)\right)$,
\begin{equation}
\begin{split}
&\det\frac{\partial\left(\Im{\boldsymbol\psi}_{\perp}(0){\rm e}^{{\rm i}\alpha}-{\bf p}_{\perp},\Re{\boldsymbol\psi}_{\perp}(t){\rm e}^{-{\rm i}\alpha}-{\bf q}_{\perp},\theta\right)}{\partial\left({\bf q}_{\perp},\Im{\boldsymbol\psi}_{\perp}\left(0\right),N_{\gamma}\right)}= \\
&\left(\cos\alpha\right)^{L-2}\det\frac{\partial\left(\Im{\boldsymbol\psi}_{\perp}(0){\rm e}^{{\rm i}\alpha}-{\bf p}_{\perp},\Re{\boldsymbol\psi}_{\perp}(t){\rm e}^{-{\rm i}\alpha}-{\bf q}_{\perp},\theta\right)}{\partial\left({\bf q}_{\perp},\Im{\boldsymbol\psi}_{\perp}\left(0\right){\rm e}^{{\rm i}\alpha},N_{\gamma}\right)}.
\end{split}
\end{equation}
Thus,
\begin{equation}
\begin{split}
&\frac{\det\left(\begin{array}{cc}
\frac{\partial\Im{\boldsymbol\psi}_{\perp}(0)}{\partial{\bf q}_{\perp}}+\frac{\sin\alpha}{\cos\alpha} & \frac{\partial\Im{\boldsymbol\psi}_{\perp}(0)}{\partial\bf p}-\frac{1}{\cos\alpha} \\
\frac{\partial\Re{\boldsymbol\psi}(t)}{\partial{\bf q}_{\perp}}-\frac{1}{\cos\alpha} & \frac{\partial\Re{\boldsymbol\psi}(t)}{\partial{\bf p}}+\frac{\sin\alpha}{\cos\alpha}
\end{array}\right)}{\det\frac{\partial\left(\Im{\boldsymbol\psi}(0),t\right)}{\partial\left({\bf p},E\right)}} \\
&=\left(\cos\alpha\right)^{-L}\left(\frac{\partial\theta}{\partial N_{\gamma}}\right)\det\left(\frac{\partial\Re\left({\boldsymbol\psi}_{\parallel}\left(t\right){\rm e}^{-{\rm i}\alpha}\right)}{\partial\left(t,\theta\right)}\frac{\partial\left(E,N_{\gamma}\right)}{\partial\Im{\boldsymbol\psi}_{\parallel}(0)}\right) \\
&\quad\times\det\left(\frac{\partial\left(\Re{\boldsymbol\psi}_{\perp}(t){\rm e}^{-{\rm i}\alpha}-{\bf q}_{\perp},{\bf p}_{\perp}-\Im{\boldsymbol\psi}_{\perp}(0){\rm e}^{{\rm i}\alpha}\right)}{\partial\left({\bf q}_{\perp},\Im{\boldsymbol\psi}_{\perp}\left(0\right){\rm e}^{{\rm i}\alpha}\right)}\right) \\
&=-\left(\cos\alpha\right)^{-L}\frac{4b^4}{\hbar}\left(\frac{\partial\theta}{\partial N_{\gamma}}\right)\det\left(\frac{\partial\Re\left({\boldsymbol\psi}_{\parallel}\left(t\right){\rm e}^{-{\rm i}\alpha}\right)}{\partial\left(t,\theta\right)}\right)^{2} \\
&\quad\times\det\left(\frac{\partial\left(\Re{\boldsymbol\psi}_{\perp}(t){\rm e}^{-{\rm i}\alpha}-{\bf q}_{\perp},{\bf p}_{\perp}-\Im{\boldsymbol\psi}_{\perp}(0){\rm e}^{{\rm i}\alpha}\right)}{\partial\left({\bf q}_{\perp},\Im{\boldsymbol\psi}_{\perp}\left(0\right){\rm e}^{{\rm i}\alpha}\right)}\right).
\end{split}
\end{equation}
The additional minus sign in the last step results from the fact that
\begin{equation}
\frac{\partial\Re\left({\boldsymbol\psi}_{\parallel}\left(t\right){\rm e}^{-{\rm i}\alpha}\right)}{\partial\theta}=-\Im{\boldsymbol\psi}_{\parallel}(0)=-\frac{1}{2}\left(\frac{\partial{N_{\gamma}}}{\partial\Im{\boldsymbol\psi}_{\parallel}(0)}\right)^{\rm T}.
\end{equation}
The remaining prefactors stem from the special form of the equations of motion, Eq.~(\ref{eq:eom}), together with the relation between $\boldsymbol\psi$ and ${\bf q},{\bf p}$, Eq.~(\ref{eq:pseudoperiodicity_orig_with_pq}).
\section{Uniqueness of the phase difference of pseudo-periodic orbits}
\label{app:uniqueness_phase}
In this appendix, we show that if $\gamma$ is quasi-periodic with phase difference $\alpha$, {\it i.e.}~
\begin{equation}
{\boldsymbol\psi}(T_\gamma)={\boldsymbol\psi}(0)\exp(-{\rm i}\alpha),
\end{equation}
with pseudo-period
\begin{equation}
T_{\rm ppo}=T_{\gamma}/m,\qquad m\in\mathbb{N},
\end{equation}
then there is no $\beta$, which is not an integer multiple of $\alpha/m$, such that for some time $t^\ast$ the trajectory satisfies
\begin{equation}
{\boldsymbol\psi}(T^\ast)={\boldsymbol\psi}(0)\exp(-{\rm i}\beta).
\end{equation}
To see this, assume there would be such a phase $\beta$ and time $T^\ast<T_{\gamma}$. Since $\beta$ is not an integer multiple of $\alpha/m$, $T^\ast$ also is not an integer multiple of $T_{\rm ppo}$ Then there is also an integer
\begin{equation}
m^\ast T^\ast<t_{\gamma}<(m^\ast+1)T^\ast
\end{equation}
and the trajectory would have to satisfy
\begin{equation}
{\boldsymbol\psi}(T_\gamma)={\boldsymbol\psi}(m^\ast T^\ast)\exp(-{\rm i}(\alpha-m^\ast\beta)),
\end{equation}
{\it i.e.}~the trajectory would be pseudo-periodic with phase difference $\alpha-m^\ast\beta$ and primitive period $T_{\gamma}-m^\ast T^\ast<T^\ast$, which is not an integer multiple of $T_{\rm ppo}$. Now we can replace $\beta$ by $\alpha-m^\ast\beta$ and repeat the argumentation again yielding an even smaller primitive period $T_{\gamma}-m^\ast T^\ast$. Since $m$ is supposed to be the largest possible number such that
\begin{equation}
{\boldsymbol\psi}(T_{\gamma}/m)={\boldsymbol\psi}(0)\exp(-{\rm i}\alpha/m)
\end{equation}
this procedure can be repeated infinitely often yielding finally $T^\ast=0$. However, having a pseudo-period $T^\ast=0$ means, that $\boldsymbol\psi(t)$ is an eigenstate of the mean-field Hamiltonian, and therefore an orbit of zero length, which are not included in the oscillatory part. In the same way, one can show that the case $T^\ast>T_{\gamma}$ leads to the same contradiction by simply replacing the roles of $\alpha$ and $\beta$. Thus, Eq.~(\ref{eq:parallel_integrals}) is indeed the correct result.
\end{appendix}
%
\bibliography{traceformula}
\bibliographystyle{apsrev4-1}
\end{document}